\documentclass[aps,amsmath]{revtex4}
\usepackage{epsfig}

\newcommand{\be}{\begin{equation}}
\newcommand{\ee[1]}{\label{#1}\end{equation}}
\newcommand{\bea}{\begin{eqnarray}}
\newcommand{\eea[1]}{\label{#1}\end{eqnarray}}
\newcommand{\bd}{\begin{displaymath}}
\newcommand{\ed}{\end{displaymath}}
\newcommand{\non}{\nonumber \\}
\begin{document}
\title{Phase-Space Description of Momentum Spectra in 
Relativistic Heavy-Ion Collisions}
\author{Werner Deinet}
\email{deinet@th.physik.uni-frankfurt.de}
\affiliation{Institut f\"ur Theoretische Physik, Johann Wolfgang
Goethe--Universit\"at, Max-von-Laue--Str.\ 1, D-60438 Frankfurt am
Main, Germany}
\author{Dirk H.\ Rischke}
\email{drischke@th.physik.uni-frankfurt.de}
\affiliation{Institut f\"ur Theoretische Physik, Johann Wolfgang
Goethe--Universit\"at, Max-von-Laue--Str.\ 1, D-60438 Frankfurt am
Main, Germany}
\date{\today}
\begin{abstract}
We consider a phase-space model for particle production 
in nuclear collisions. Once the multiplicities 
of the individual particle species are known, 
single-inclusive momentum spectra can be computed 
after making simplifying assumptions for the
matrix element for multiparticle production. Comparison of the
calculated spectra with data for pions and kaons from central Pb+Pb collisions 
at $E_{\rm Lab}=158$ AGeV reveals a residual longitudinal
phase-space dominance in the final state of the reaction. 
We account for this by modifying
the isotropic, relativistic invariant phase space
in a way which retains boost invariance in beam
direction but suppresses large transverse momenta. Adjusting a
single parameter, we obtain a reasonably good description of
transverse momentum and rapidity spectra for both pions and kaons.
\end{abstract}
\maketitle
\section{Introduction}

In relativistic heavy-ion collisions one can study strongly
interacting matter under extreme conditions \cite{QM2004}.
Of particular interest is the question whether a novel
phase of nuclear matter, the so-called quark-gluon plasma (QGP),
can be created in such collisions. In this phase,
quarks and gluons are deconfined and chiral symmetry is
restored. Lattice QCD calculations \cite{latt} predict that, in
thermodynamical equilibrium, this phase occurs at temperatures
above $\sim 170$ MeV (for vanishing baryon chemical potential).
Such temperatures are expected to be reached and exceeded for collisions
at CERN-SPS and BNL-RHIC energies, respectively.

In nucleus-nucleus collisions at ultrarelativistic energies,
secondary particles are copiously produced over a small region
in space-time. It is then natural to assume that interactions between 
these particles drive the system towards (local) 
thermodynamical equilibrium. Indeed, 
measured single-inclusive transverse momentum spectra
of individual particle species show an exponential decrease 
similar to that expected in thermal equilibrium.
Following this line of arguments, 
one has studied the evolution of the system using models
based on ideal fluid dynamics
where the system is assumed to be in local thermodynamical
equilibrium. At RHIC energies, such models are able to reproduce the data
for single-inclusive transverse momentum spectra of various
particle species \cite{huovinen}. A particular success of the
fluiddynamical approach was the correct prediction of the magnitude
of elliptic flow.
The fact that data on elliptic flow seem to saturate the
ideal fluid limit has recently led to the speculation that matter
created in collisions at RHIC energies is a so-called
{\em strongly coupled QGP\/} with very small viscosity over entropy
ratio \cite{sQGP}.

However, the fluiddynamical approach is not without problems.
A long-standing problem is the so-called ``HBT puzzle'', i.e.,
the fact that HBT correlation radii and single-inclusive
particle spectra cannot be simultaneously described
within the fluiddynamical approach \cite{kolb}.

Another problem of the fluiddynamical approach
concerns the reproduction of the measured particle ratios.
Under the assumption of local thermodynamical equilibrium,
as well as conservation of electric charge and strangeness,
one can determine from such particle ratios the temperature and the 
baryon chemical potential of the system on the hypersurface
where chemical freeze-out occurs, i.e., where inelastic
collisions cease and the multiplicities
of produced particles no longer change \cite{Braun,Becca,Cley}. 
The values for the chemical freeze-out temperature and baryon chemical
potential thus obtained indicate that the system undergoes 
chemical freeze-out well before thermal freeze-out, i.e., when also 
elastic collisions cease and
the single-inclusive particle spectra no longer change their shape.
If one accounts for this in the fluiddynamical evolution
by allowing for departures from chemical equilibrium \cite{hirano}
and thus keeps the particle ratios at their experimental values,
one is no longer able to describe data for collective flow
\cite{gyul}. (An exception is the study of Ref.\ \cite{rasanen} which 
reproduces the transverse mass spectra {\em and\/} 
multiplicities for pions, kaons, and (anti-)protons for collisions
at RHIC energies assuming simultaneous thermal and chemical freeze-out.)

In this paper, we pursue a different strategy of describing
single-inclusive momentum spectra of particles produced
in heavy-ion collisions.
Our arguments are solely based on the phase space of the
particles in the final state.
Thus, we follow a well-known and well-documented
approach in the physics of hadron-hadron collisions \cite{Hag,Byck,Hama}
and apply it to nucleus-nucleus collisions 
\cite{knoll,Risch,hsu,Koch,Becca2}.
In order to determine the phase space, we need to know the
experimentally measured multiplicities of secondary particles. 
We also have to make simplifying assumption for the
matrix element of multiparticle production. We can then
determine single-inclusive particle spectra and compare them with the
experimentally measured spectra.

This phase-space model to describe single-inclusive particle
spectra represents an approach with the minimum number of
assumptions concerning the dynamics of the reaction.
In its most simple form it relies
solely on the constraints imposed by the 
kinematics of the reaction, such as total energy and particle
multiplicities.
Deviations from the phase-space model are a clear indication
for nontrivial dynamics beyond pure kinematics.

Our approach is complementary to the
fluid-/thermodynamical approach. There, the exponential decay
of transverse mass spectra as seen in the data is the motivation 
to argue that matter
is in local thermodynamical equilibrium. On the other hand, 
the data deviates from this exponential
decay, especially in the region of small transverse momenta.
These deviations are commonly attributed to resonance decays
and to collective motion.
Particle ratios can be predicted, once temperature and baryon chemical
potential at chemical freeze-out are known.
On the other hand, in our approach particle multiplicities serve
as input to determine the size of multiparticle phase space.
Once they are known, single-inclusive particle spectra can be
computed. As we shall see, the calculated 
transverse mass spectra also decay exponentially. 
Similar to the fluid-/thermodynamical approach
we then interpret the deviation of the data from a pure exponential decay
as a signature for nontrivial dynamics, such as resonance decays and
collective flow of matter.

For this study, as an example we use data 
\cite{NA49,Afan,Misch,NA49lam,NA49xi,NA49ome} of the NA49 collaboration 
for the 5\% most central Pb+Pb collisions at a beam energy of $E_{\rm Lab}
= 158$ AGeV, corresponding to $\sqrt{s_{NN}} = 17.3$ AGeV in
the center-of-momentum frame. 
For the 5\% most central collisions, the average number of 
wounded nucleons is $\langle N_w \rangle=362$.
Therefore, the total energy deposited in the reaction is estimated 
to be
$\sqrt{s}=\sqrt{s_{NN}}\frac{\langle N_w \rangle}{2}=3131.3$ GeV.
In principle, we could also have used data from other experiments,
but the advantage of the NA49 experiment is that it provides
full rapidity coverage. It thus allows us to investigate
rapidity spectra in addition to the transverse mass spectra.

This paper is organized as follows. In Sec.\ \ref{II} we define and
compute the phase-space integral. In practice, 
the computation is not possible unless one makes simplifying 
assumptions for the matrix element for multiparticle production.
We investigate a product ansatz for the matrix element, which allows
to factorize the phase-space integral and thus facilitates the
computation. We then study two special cases in more detail, 
one which is completely isotropic
in phase space and preserves general
Lorentz invariance, and another which incorporates the
longitudinal phase-space dominance of the reaction and only
preserves Lorentz invariance under boosts in beam direction.
In Sec.\ \ref{III} we compute single-inclusive momentum spectra,
i.e., transverse mass and rapidity spectra,
from the phase-space integral and point out differences to 
momentum spectra in thermal equilibrium.
In Sec.\ \ref{IV} we confront the calculated spectra with
experimental data. To this end, it was also necessary
to reconstruct the multiplicities
of unmeasured particle species using symmetry considerations,
an empirical rule, and a simple quark model. 
Section \ref{V} concludes this work with a summary and an outlook.
Our units are $\hbar=c=k_B=1$ and the metric tensor is chosen as
$g^{\mu \nu} = (+,-,-,-)$.

\section{The phase-space integral} \label{II}

Let us consider the cross section for production 
of secondaries in collisions of elementary particles (or
nuclei).
The particles (or nuclei) in the incoming channel have
masses $m_a,\, m_b$ and four-momenta $p_a,\, p_b$. The 
center-of-momentum energy (squared) of the reaction is 
$s \equiv P^2$, where $P^\mu = (E,\mathbf{P}) = p_a^\mu + p_b^\mu$ is the 
total four-momentum.
We assume that there are $n$ particles in 
the final state with masses $m_1, \ldots m_n$ and four-momenta
$p_1^\mu, \ldots p_n^\mu$, 
where $p_i^\mu=(E_i, {\bf p}_i)$, with $E_i = \sqrt{{\bf
p}_i^2 +m_i^2}$. The cross section $\sigma$ for this reaction is given by
\be  
\sigma=\frac{1}{F}\, I_n(P^\mu)\;,
\ee[eqsig]
where 
\be   
F \equiv 2\,\sqrt{\lambda(s,m_a,m_b)}\,(2\pi)^{3n-4}   
\ee[1]
is the flux factor, with
$\lambda(x,y,z)=x^2+y^2+z^2-2xy-2yz-2zx$ being the kinematical function 
\cite{Byck}. 
$I_n$ is an integral over the $3n$-dimensional invariant momentum space of 
the secondaries,
\be  
I_n(P^\mu)=\int{\prod^n_{i=1}\frac{d^3p_i}{2E_i}\,
\delta^4(P^\mu-\sum_j{p_j^\mu})\, T({\bf p}_i)}\;. 
\ee[eqIn]  
The integrand consists of a
four-dimensional, four-momentum conserving $\delta$-function 
and of the square $T(\mathbf{p_i})$ of the transition matrix element.

The secondaries produced in a reaction come in different
species $r$. The multiplicity of particles of species $r$ is denoted
by $n_r$, such that $n = \sum_r n_r$. In order to compute 
$I_n$ one needs to specify the multiplicities $n_r$.
These differ from event to event. Usually,
event-averaged quantities are given in the literature. We shall 
exclusively use these event-averaged
multiplicities in the following.
We shall see that our results do not change qualitatively
if we vary the multiplicities $n_r$ within 
a factor of $3 \sqrt{n_r}$, where we assume $\sqrt{n_r}$ to be a measure
for the statistical error.
One could also study multiparticle production on an event-by-event
basis  \cite{werner} in order to examine particle number fluctuations,
but this is beyond the scope of the present work.

In order to calculate $I_n$, 
according to Eq.\ (\ref{eqIn}) $3n$ integrations have to be
performed, where in our case $n$ is larger than 2000. 
The $\delta^4$-function prevents that these integrals
decouple. Therefore, the computation of $I_n$ 
requires in general numerical methods.
One can distinguish three approaches, recursion relations
\cite{Hag,Byck,Fri}, Monte Carlo methods 
\cite{Jam,Kleiss}, and the statistical approach \cite{Byck}. 
In this work we employ the latter, because they are most appropriate
for a large number of secondaries. 
The essential idea is to use standard methods of statistical mechanics 
\cite{Lurc,Krzy,Krzy1,Bial,Kaja,Kaja1}. In
statistical mechanics, calculations performed in the 
microcanonical ensemble are difficult, because the energy is fixed. 
Therefore, one replaces the microcanonical ensemble by the 
canonical ensemble, where the energy can vary, but the 
temperature is fixed. In the thermodynamical limit, i.e.,
for large particle numbers, both approaches yield identical
results. The same idea is applied here to compute $I_n$,
but there exist several methods which differ in detail.
Some use the central-limit theorem of probability theory, as suggested
and explored by Khinchin \cite{Khin}. 
In this work, we employ the saddle-point method \cite{Kaja},
which yields, in the approximation used here, the same result as the
methods based on the central-limit theorem.

\subsection{The calculation of the phase-space integral}

In general, for multiparticle reactions the matrix element $T({\bf
p}_i)$ is too complicated to calculate analytically. Therefore, in
order to make progress one resorts to simple approximations for
$T({\bf p}_i)$. Such an approximation can be motivated by the symmetries of
the problem, and it should not spoil the Lorentz invariance of $I_n$. 
Our particular ansatz used in the following is
\be
T({\bf p}_i) = \prod_{i=1}^n f_i({\bf p}_{\perp i})\;.
\ee[Tpi]
The most simple case, without any assumptions about the symmetries
of the problem and preserving full Lorentz invariance, is
\be
f_i({\bf p}_{\perp i}) = 1\;.
\ee[FLI]
Another case studied here is motivated by the longitudinal phase-space
dominance in hadron-hadron and high-energy nuclear collisions.
Choosing
\be 
f_i({\bf p}_{\perp i})=\exp \left[-a (m_{\perp i}-m_i)\right]\;,
\ee[eqf]
where $m_{\perp i} = \sqrt{p_{\perp i}^2 + m_i^2}$ is
the transverse mass,
the transverse phase space is exponentially suppressed.
This ansatz preserves boost invariance with respect to the beam axis,
but is no longer invariant under Lorentz transformations perpendicular
to this direction.

Inserting the ansatz (\ref{Tpi}) into Eq.\ (\ref{eqIn}), we
obtain
\be  
I_n(P^\mu)=\int{\prod^n_{i=1}\frac{d^3p_i}{2E_i}\, 
f_i({\bf p}_{\perp i})\,\delta^4(P^\mu-\sum_j{p_j^\mu})}\;. 
\ee[eqInf]  
The four-momentum conserving delta-function introduces a mutual dependence
of the various momentum integrals and thus makes the integral $I_n$
difficult to compute directly. This problem is similar to that which 
one encounters in the computation of the microcanonical
partition function in statistical mechanics. In that case,
the computation is facilitated by considering
the canonical partition function as the Laplace transform
of the microcanonical one. Technically, the Laplace transformation 
decouples the momentum integrals. Once the 
canonical partition function is known, an inverse
Laplace transformation brings us back to microcanonical partition
function which we wanted to compute originally.
In our case, the analogue of the canonical partition function
in statistical mechanics is the (four-dimensional) Laplace transform 
of the integral $I_n$,
\be 
\Phi_n(\alpha_\mu)= \int d^4P \, \exp \left(-\alpha_\nu P^\nu \right)
\, I_n(P^\mu)\;. 
\ee[eqPhi]
The Laplace transform $\Phi_n$ is a function of the 
four-vector $\alpha_\mu$. 
Inserting $I_n$ into Eq.\ (\ref{eqPhi}) and performing
the integration over $P^\mu$, we obtain
\bea 
\Phi_n(\alpha_\mu)& =&\int{d^4P\prod_{i=1}^n \frac{d^3p_i}{2E_i}\,
f_i({\bf p}_{\perp i}) \, \delta^4(P^\mu-\sum_j{p_j^\mu}) 
\, \exp \left(-\alpha_\nu P^\nu \right)}  \nonumber \\
& =& \int{\prod_{i=1}^n \frac{d^3p_i}{2E_i}\, 
f_i({\bf p}_{\perp i})\, \exp \left(-\alpha_\nu p_i^\nu\right)}  \;.
\eea[eqPhi4]
We define the functions
\be 
\phi_i(\alpha_\mu)=\int\frac{d^3p_i}{2E_i}\,f_i({\bf p}_{\perp i})\,
\exp \left(-\alpha_\nu p_i^\nu \right)\; ,   
\ee[eqphi] 
such that 
\be  
\Phi_n(\alpha_\mu)=\prod_{i=1}^n\phi_i(\alpha_\mu)   \;.
\ee[eqPhi2] 
The original integral $I_n$ can be obtained from the inverse
Laplace transformation,
\be 
I_n(P^\mu)=\frac{1}{(2\pi i)^4} \int_{c_\mu-i\infty}^{c_\mu+i\infty}
d^4\alpha\, \exp \left(\alpha_\nu P^\nu\right)\, \Phi_n(\alpha_\mu)\;.  
\ee[eqIn1]
The path of integration is parallel to the imaginary axis, where
$c_\mu$ is a constant 4-vector. Each component $c_\mu$ 
is chosen such that the path of integration lies in a range
of values for $\alpha_\mu$, where the integrand does 
not have any singularities.

We proceed by taking $\Phi_n$ into the argument of the exponential function,
\be 
I_n(P^\mu)=\frac{1}{(2\pi i)^4} \int_{c_\mu -i\infty}^{c_\mu
+i\infty} d^4\alpha \, \exp \left[F(\alpha_\mu)\right] \;,
\ee[eqIn3]
where we defined
\be 
F(\alpha_\mu)\equiv \alpha_\nu P^\nu + \ln\Phi_n(\alpha_\mu) \;.
\ee[eqF]
In general, the integral (\ref{eqIn3}) cannot be solved analytically. 
Statistical methods are used to obtain an approximative solution. Here,
we employ the saddle-point approximation \cite{Byck,Kaja}.
The saddle point of the integrand, $\beta_\mu$, is given by the condition
\be
\left.
\frac{\partial F(\alpha_\mu)}{\partial \alpha_\nu} \right|_{\alpha_\mu =
\beta_\mu} =0\;,
\ee[spc]
or, with Eq.\ (\ref{eqF}),
\be
P^\nu = - 
 \frac{\partial \ln \Phi_n(\beta_\mu)}{\partial \beta_\nu}
= - \sum_{i=1}^n \frac{1}{\phi_i(\beta_\mu)}
\int\frac{d^3p_i}{2E_i}\,f_i({\bf p}_{\perp i})\, p_i^\nu
\exp \left(-\beta_\lambda p_i^\lambda \right)
 \;.
\ee[2]
Note that, in the statistical mechanical problem of computing
the microcanonical partition function, the 
parameter $\beta_\mu$ would correspond to the inverse temperature.
Going into the rest frame of the saddle point, $\beta^\mu = (\beta,
{\bf 0})$ and using the fact that $f({\bf p}_{\perp i})$ and
$p_i^0 \equiv E_i$ are even functions of momentum, we
observe that this rest frame is identical to the
center-of-momentum frame, $P^\mu = (\sqrt{s}, {\bf 0})$.
Thus, $\beta^\mu$ and $P^\mu$ must be proportional to each other, 
$P^\mu \sim \beta^\mu$, in any
frame which differs from the rest frame by a boost in longitudinal 
direction. 
We now determine the constant of proportionality.
To this end, we compute the four-gradient of $\ln \Phi_n$
at the saddle point assuming $\Phi_n$ to be a function of the
absolute value $\beta = \sqrt{\beta_\nu \beta^\nu}$ of $\beta^\mu$,
and its direction, $\hat{\beta}^\mu \equiv \beta^\mu/\beta$.
Then,
\be
P^\nu = -  \frac{\partial \ln \Phi_n(\beta, \hat{\beta}_\mu)}{\partial 
\beta_\nu} 
= - \frac{\partial \ln \Phi_n(\beta, \hat{\beta}_\mu)}{\partial 
\beta} \, \hat{\beta}^\nu
- \frac{\partial \ln \Phi_n(\beta, \hat{\beta}_\mu)}{\partial 
\hat{\beta}_\lambda}\, 
\frac{1}{\beta}\, {B_\lambda}^\nu\;,
\ee[3]
where
\be
{B_\lambda}^\nu \equiv \beta\, 
\frac{\partial \hat{\beta}_\lambda}{\partial \beta_\nu}
= {g_\lambda}^\nu - \hat{\beta}_\lambda \hat{\beta}^\nu
\ee[B]
is the projector onto the direction which is four-transverse
to $\beta^\mu$. However, since $\beta^\mu$ and
$P^\mu$ are proportional to each other, 
the part proportional to ${B_\lambda}^\nu$ must be
zero, i.e., there is no dependence of $\ln \Phi_n$ on 
$\hat{\beta}^\mu$, and we have
\be
P^\nu = -  \frac{\partial \ln \Phi_n(\beta)}{\partial \beta}
\, \hat{\beta}^\nu\;.
\ee[4]
Going to the center-of-momentum frame, this equation serves
to (implicitly) determine the absolute value of $\beta$,
\be
\sqrt{s} = - \frac{\partial \ln \Phi_n(\beta)}{\partial \beta} 
= \frac{1}{2} \sum_{i=1}^n \frac{1}{\phi_i(\beta)}
\int d^3p_i\,f_i({\bf p}_{\perp i})\, e^{-\beta E_i}\;.
\ee[sqrts]
This equation relates $\beta$
to the center-of-momentum energy
$\sqrt{s}$. In the analogous problem of statistical mechanics,
such a saddle-point condition
relates the inverse temperature (the independent thermodynamical 
variable in the canonical ensemble) to the
total energy (the independent thermodynamical variable in the
microcanonical ensemble).

We now proceed with the evaluation of the integral (\ref{eqIn3}).
We choose the integration path to lead through the saddle point
and change the integration
variable to $t^\mu \equiv -i (\alpha^\mu - \beta^\mu)$.
For large particle numbers the integrand is sharply peaked
around the saddle point, $t^\mu = 0$, and we may employ a Taylor expansion of
the function $F(\alpha_\mu)$.
For our numerical calculations presented below, we checked that
terms higher than second order in this expansion are at most of the order of 
$\sim 10^{-4}$ and thus insignificant. We therefore restrict
the following consideration to terms of second order in $t^\mu$, 
\be 
F(\alpha_\mu) = F(\beta_\mu) 
- \frac{1}{2}\, t^\nu t^\lambda \,F_{\nu\lambda} +O(t^3)\;,
\ee[eqF4]
where we used the saddle-point condition (\ref{spc}) to
eliminate the linear term in $t^\mu$, and where 
the second derivative of $F(\alpha_\mu)$ with respect to
$\alpha_\nu$ at the saddle point is denoted as
\be
F_{\nu \lambda} \equiv \left.
\frac{\partial^2 F(\alpha_\mu)}{\partial \alpha^{\nu}
\partial \alpha^{\lambda}}
\right|_{\alpha_\mu = \beta_\mu}\,.
\ee[5]
Using Eq.\ (\ref{eqF}) and the fact that
$\ln \Phi_n$ only depends on the absolute value of $\beta^\mu$,
we can further evaluate $F_{\nu \lambda}$,
\be
F_{\nu \lambda}
= \frac{\partial^2 \ln \Phi_n(\beta)}{\partial \beta^{\nu}
\partial \beta^{\lambda}}
= \frac{\kappa}{\beta^2}\, \hat{\beta}_\nu \hat{\beta}_\lambda
- \frac{\sqrt{s}}{\beta}\, B_{\nu \lambda}\;,
\ee[6]
where we employed Eqs.\ (\ref{B}) and (\ref{sqrts}) and defined
\be
\kappa \equiv \beta^2 \,
\frac{\partial^2 \ln \Phi_n(\beta)}{\partial \beta^2}
=  \frac{\beta^2}{4} \sum_{i=1}^n \frac{1}{\phi_i(\beta)}
\left\{ 2\int d^3p_i\,f_i({\bf p}_{\perp i})\, E_i\, e^{-\beta E_i}
- \frac{1}{\phi_i(\beta)} \left[ \int d^3 p_i \, f_i({\bf p}_{\perp i})\,
e^{- \beta E_i} \right]^2 \right\}\;.
\ee[7]
For the explicit expression on the right-hand side we have chosen
the rest frame of $\beta^\mu$. Inserting the Taylor expansion into
Eq.\ (\ref{eqIn3})  we obtain 
\be 
I_n(P^\mu) \simeq \frac{e^{\beta \sqrt{s}} \Phi_n(\beta)}{(2\pi)^4}
\int d^4t \, \exp \left\{-\frac{1}{2 \beta^2}\,\left[
\kappa \,(t^\nu \hat{\beta}_\nu)^2
- \beta \sqrt{s}\, t^\nu B_{\nu \lambda} t^\lambda \right] \right\}
\; .
\ee[eqIn4]
We evaluate the $t$-integral in the rest frame of $\beta^\mu$,
where
\be
\kappa \,(t^\nu \hat{\beta}_\nu)^2
- \beta \sqrt{s}\, t^\nu B_{\nu \lambda} t^\lambda =
\kappa\, t_0^2 + \beta \sqrt{s}\; {\bf t} \cdot {\bf t}\;,
\ee[kappa]
i.e., the integral decouples into four Gaussian integrals. The final
result is
\be 
I_n(P^\mu)
\simeq \frac{e^{\beta\sqrt{s}}}{4\pi^2}\, \Phi_n(\beta)\,
\frac{\beta}{\sqrt{\kappa}}\, 
\left(\frac{\beta}{\sqrt{s}}\right)^{3/2}\;.
\ee[eqRn0]
For large multiplicities $n$, the logarithm of this expression
can be estimated as
\be
\ln I_n(P^\mu) \simeq \beta \sqrt{s} + \ln \Phi_n(\beta)\;.
\ee[8]
Here, we have only kept terms of order $O(n)$, such as $\sqrt{s}$.
Terms proportional to $\ln \beta$ are of order $O(1)$, 
because $\beta$ itself is $\sim O(1)$, 
see e.g.\ Eq.\ (\ref{spm=0}) below for the massless case.
Also, terms of order $\ln \sqrt{s}$ and $\ln \kappa$ are of order
$O(1)$, because $\sqrt{s}$ and $\kappa$ are of order $O(n)$, 
see e.g.\ Eq.\ (\ref{kapm=0}) below, and thus
$\ln \sqrt{s} \sim \ln \kappa \sim O(1)$.
Thus, with Eq.\ (\ref{sqrts}) we obtain
\be
\frac{\partial \ln I_n(P^\mu)}{\partial \sqrt{s}}
\equiv 2 \sqrt{s}\, \frac{\partial \ln I_n(P^\mu)}{\partial s}
\simeq \beta + \frac{\partial \beta}{\partial \sqrt{s}}
\left( \sqrt{s} + \frac{\ln \Phi_n(\beta)}{\partial \beta} \right)
\equiv \beta\;.
\ee[20]

\subsection{Special cases}

We now discuss some special cases of the result (\ref{eqRn0}).
In the case (\ref{FLI}),
$\Phi_n$ is invariant
under any Lorentz transformation and thus can only depend on
the Lorentz-scalar quantity $\beta= \sqrt{\beta^\nu \beta_\nu}$.
It may be computed in any frame, but the rest frame of $\beta^\mu$
is particularly convenient. In this frame \cite{Grad,Abra},
\be 
\phi_i(\beta)=2\pi \int_0^\infty d p_i \, \frac{p_i^2}{E_i}\,
e^{-\beta E_i}
=\frac{2\pi m_i}{\beta} K_1(\beta m_i)\;, 
\ee[eqphi1] 
and $\Phi(\beta)$ is readily obtained from Eq.\ (\ref{eqPhi2}).
The saddle-point condition (\ref{sqrts}) reads
\be  
\sqrt{s}=\frac{2n}{\beta}+\sum_{i=1}^n
m_i{\frac{K_0(\beta m_i)}{K_1(\beta m_i)}} \;,
\ee[eqws1]
and $\kappa$ is determined as
\be
\kappa = 4n - \beta \sqrt{s} + \beta^2 \sum_i m_i^2 \left\{ 1-
\left[ \frac{K_0(\beta m_i)}{K_1(\beta m_i)} \right]^2 \right\}\;.
\ee[9]
If all particles are massless, we obtain
\be
\phi_i(\beta) = \frac{2 \pi}{\beta^2}\;,
\ee[phim=0]
where the saddle point is given by
\be
\beta = \frac{2n}{\sqrt{s}}\;.
\ee[spm=0]
Finally, 
\be
\kappa = 4n - \beta \sqrt{s} \equiv 2n\;,
\ee[kapm=0]
such that 
\be
I_n(P^\mu) \simeq \frac{e^{2n} \pi^{n-2}}{2^n\, n^2}  \, 
\left( \frac{\sqrt{s}}{n} \right)^{2n-4}\;.
\ee[appresult]
The exact result is \cite{Kaja}
\be
I_n(P^\mu) = \frac{\pi^{n-1}}{2^{n-1} (n-1)! (n-2)!}\,
\sqrt{s}^{2n-4}\;.
\ee[exaresult]
For large $n$, one may approximate the factorials 
with the help of Stirling's formula. Using $(1-x/n)^n \rightarrow
e^{-x}$ for $n \rightarrow \infty$, 
one then observes that the result
(\ref{appresult}) obtained via the saddle-point method
is consistent with the exact result (\ref{exaresult}).

\begin{center}
\begin{figure}
\epsfig{file=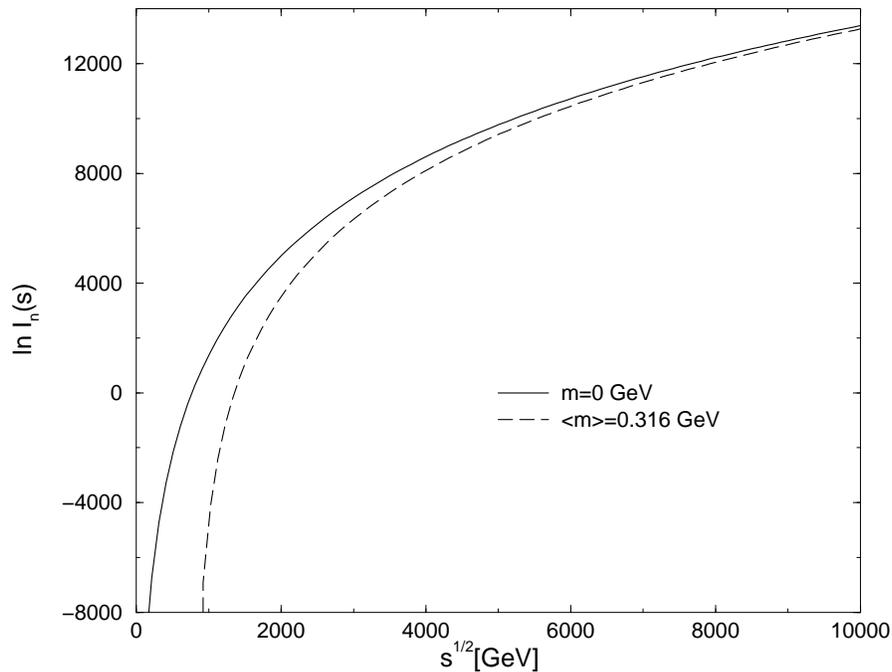,width=9cm,angle=-90}
\caption{\small The logarithm of the phase-space integral for
$n=2611$ massless particles (solid curve) and for
massive particles as listed in Table \ref{tab3} (dashed curve).
Both curves are computed for $a=0$.}
\label{fig4}
\end{figure}
\end{center}

In the case (\ref{eqf}), $\Phi_n$ is invariant
under Lorentz transformations along the $z$-axis.
Again, we compute it in the rest frame of $\beta^\mu$.
Defining the function
\be
{\cal J}_{l,m}(x,y)
\equiv 2 \pi \, e^y \int_x^\infty dz\, z^l \, e^{- y z/x} \, K_m(z)\;,
\ee[10]
we obtain
\be  
\phi_i(\beta) = \beta^{-2}\, {\cal J}_{1,0}(\beta m_i, a m_i)\;.
\ee[eqphi4]
The saddle-point condition (\ref{sqrts}) now reads
\be
\sqrt{s} =  \frac{1}{\beta} \sum_{i=1}^n \frac{{\cal J}_{2,1}(\beta
m_i,a m_i)}{{\cal J}_{1,0}(\beta m_i,a m_i)}\;.
\ee[11]
The coefficient $\kappa$ is determined as
\be
\kappa =  \sum_{i=1}^n 
\left\{ \frac{{\cal J}_{3,0}(\beta m_i,am_i) + 
{\cal J}_{2,1}(\beta m_i,am_i)}{{\cal J}_{1,0}(\beta m_i,am_i)}  
 - \left[\frac{{\cal J}_{2,1}(\beta m_i,am_i)}{{\cal J}_{1,0}(\beta
m_i,am_i)} \right]^2 \right\}\;.
\ee[12]
In order to get a feeling for the magnitude of the
phase-space integral consider the hadron multiplicities given in 
Table \ref{tab3}. There are $n= 2611$ secondaries with a total rest mass
of $\sum{m_i}=824.65$ GeV. The average particle mass is 
$\langle m_i \rangle \simeq 0.316$ GeV.
The total available energy is $\sqrt{s}=3131.3$ GeV, see introduction. 
For $a=0$, this results in $I_n \simeq 0.5 \cdot
10^{2867}$, or $\ln I_n \simeq 6600$. The saddle point of the $t$-integration
is located at $\beta = 2.027\, {\rm GeV}^{-1}$.
In Fig.\ \ref{fig4} we show the logarithm of the phase-space integral 
as a function of $\sqrt{s}$ for 2611 massless particles and 
for the set of 2611 particles of Table \ref{tab3}.
For a given $\sqrt{s}$, the result for massive particles is always 
below that of massless particles, since some of the energy is
used to generate the rest mass of the particles and is no longer
available for the phase-space volume occupied by the system.
Only when $\sqrt{s}\rightarrow\infty$, the massive case
approaches the result for massless particles. 

\begin{center}      
\begin{figure}
\epsfig{file=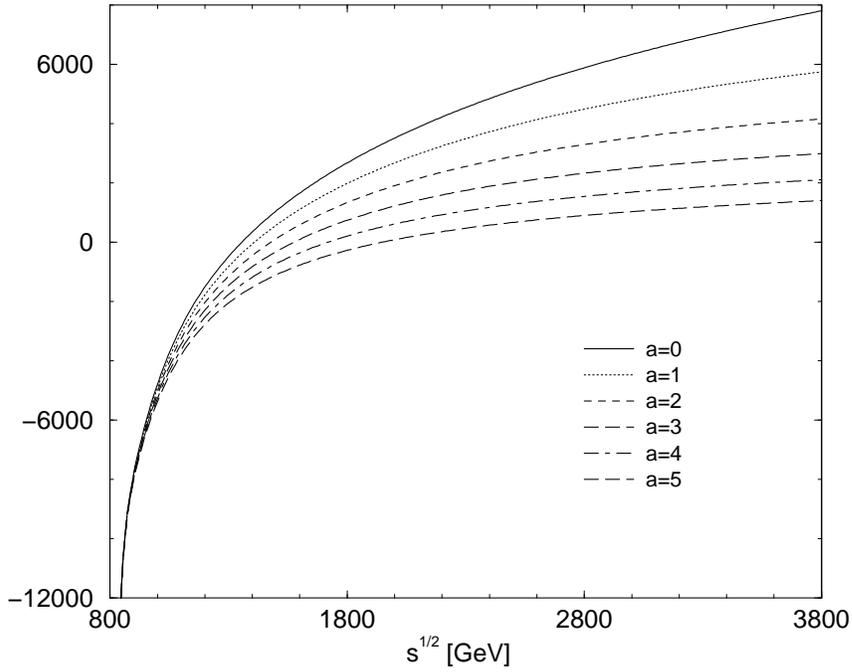,width=9cm,angle=-90}
\caption{\small The phase-space integral for particles
as listed in Table \ref{tab3} for different values of the
parameter $a$ (in units of GeV$^{-1}$).}
\label{fig21}
\end{figure}
\end{center}

The influence of the parameter $a$ on the value of the
phase-space integral is demonstrated in Fig.\ 
\ref{fig21}. The phase-space integral is 
again computed with the set of particles
listed in Table \ref{tab3}.
The value of the integral decreases with increasing $a$,
which is naturally explained by the exponential suppression
of transverse phase space, cf.\ Eq.\ (\ref{eqf}).

\section{Single-inclusive particle spectra} \label{III}

In this section we demonstrate how one can compute single-inclusive 
particle spectra from the phase-space integral discussed in the previous
section. For comparison, we also present the calculation of
single-inclusive spectra in thermal equilibrium.

\subsection{Particle spectra from phase space}

According to Eq.\ (\ref{eqsig}),
the differential cross section for inclusive production of, say,
the $n$th secondary particle is
\bea
E_n\, \frac{d^3\sigma}{d^3p_n} & = &\frac{E_n}{F}\,
\frac{d^3I_n(P^\mu)}{d^3p_n} \non
& = & \frac{f_n({\bf p}_{\perp n})}{2F} \int \prod_{i=1}^{n-1} \frac{d^3
p_i}{2E_i}\, f_i({\bf p}_{\perp i})\, \delta^4(P^\mu-p_n^\mu-\sum_j
p_j^\mu) 
\non
& = & \frac{f_n({\bf p}_{\perp n})}{2F}\, I_{n-1}(P^\mu-p_n^\mu)\;.
\eea[eqInd3p]
$I_{n-1}$ is the phase-space integral for the first $n-1$ particles
at a reduced center-of-momentum frame energy, 
\be
s' \equiv (P - p_n)^2 = s + m_n^2 - 2\, E_n\, \sqrt{s}
=s+m_n^2-2\, m_{\perp n}\, \cosh y_n\, \sqrt{s}  \;.
\ee[eqs']
The single-inclusive particle spectrum is obtained from
the differential cross section (\ref{eqInd3p}) via
\be
E_n\, \frac{d^3n_n}{d^3p_n} = 
\langle n_n \rangle\, \frac{E_n}{\sigma}\, \frac{d^3\sigma}{d^3p_n}
= \langle n_n \rangle\,
\frac{f_n({\bf p}_{\perp n})}{2}\, \frac{I_{n-1}(P^\mu-p_n^\mu)}{
I_{n}(P^\mu)}\;.
\ee[invspec]
As discussed in the previous section, 
the calculation of the phase-space integrals on the right-hand side
will be done in the rest frame of the collision.
However, because there are $n-1$ particles in the numerator and
$n$ particles in the denominator, the rest frame of
$I_{n-1}$ will be different from that of $I_n$.
This is irrelevant if the phase-space integral is Lorentz invariant,
such that $I_{n-1}(P^\mu-p_n^\mu) \equiv I_{n-1}(s')$ and
$I_n(P^\mu) \equiv I_n(s)$. If the phase-space integral is only
Lorentz invariant with respect to boosts along the $z$-axis,
this difference is also irrelevant if the $n$th particle moves along
this axis. If not, we assume the recoil in transverse direction
to be sufficiently small such that the rest frame of $I_{n-1}$
is approximately the same as that of $I_n$.
This is certainly a good approximation for large $n$.

Transverse mass spectra are obtained from Eq.\ (\ref{invspec}) 
after integrating over the transverse polar angle,
\be    
\frac{1}{m_{\perp n}}\,\frac{d^2n_n}{dm_{\perp_n} dy_n}
\simeq \pi \, \langle n_n \rangle \,
f_n({\bf p}_{\perp n})\,\frac{I_{n-1}(s')}{I_n(s)}  \;.
\ee[eqtrm1]
An additional integration over $m_{\perp n}$
yields the rapidity spectrum,
\be
\frac{dn_n}{dy_n} \simeq \pi\, \langle n_n \rangle \int_{m_n}^\infty 
dm_{\perp n}\,m_{\perp n}\, f_n({\bf p}_{\perp n})
\, \frac{I_{n-1}(s{'})}{I_n(s)}\;.  
\ee[eqdndy]
A further integration over $y_n$ then 
gives the multiplicity of the $n$-th particle, 
which is an input parameter. This integration serves as a check for
the consistency of our results.

In the following, we discuss several special cases of the general
result (\ref{eqtrm1}) for the transverse mass spectra.

\subsection{Special cases}

Let us consider the case (\ref{eqf}) for the function $f_i({\bf
p}_{\perp i})$. The other case (\ref{FLI}) can be simply obtained by
setting $a=0$ in the final result.
In Eq.\ (\ref{eqtrm1}), exponentiate the numerator $I_{n-1}(s')$ and
expand $\ln I_{n-1}(s')$ to first order in $s' - (s+m_n^2) = - 2\,
 m_{\perp n} \cosh y_n \sqrt{s}$ \cite{Fri}. This is a good 
approximation, since the energy of a single particle is
small compared to the total energy in the system,
$E_n \equiv m_{\perp n} \cosh y_n \ll \sqrt{s}$. We may also drop
$m_n^2$ compared to $s$. We obtain
\be
\frac{1}{m_{\perp n}}
\frac{d^2n_n}{dm_{\perp n}dy_n} \simeq \pi \, \langle n_n \rangle\, 
e^{a\, m_n}\, 
\frac{I_{n-1}(s)}{I_n(s)}\, \exp \left( -\frac{m_{\perp n}}{T^*} \right)\;.
\ee[eqdndmdy8]
Here we defined the slope parameter
\be
\frac{1}{T^*} \equiv 2\,  \sqrt{s}\, \cosh y_n \,
\frac{d}{ds}\ln I_{n-1}(s)  + a 
\simeq \beta\, \cosh y_n + a\;,
\ee[eqT1]
where we have used Eq.\ (\ref{20}) and the fact that $n-1 \simeq n$
for large multiplicities.
The transverse mass spectra exhibit an exponential decay in the
transverse mass $m_{\perp n}$ with a slope
parameter given by Eq.\ (\ref{eqT1}). Note that $T^*$ is
not a free fit parameter that one could adjust to describe the
data, like in global thermal fits of
single-inclusive particle spectra. 
In contrast, it is unambiguously determined by the value of 
the saddle point $\beta$, the rapidity
$y_n$ of the particle considered, and the parameter $a$.
The slope parameter depends on the mass
of the considered particle only through the position $\beta$ of 
the saddle point of the phase-space integral.
However, this dependence is very
weak, because the position of the saddle point is largely decided by
the values of the total multiplicity $n$ and the total
available energy $\sqrt{s}\gg m_n$ . 
Quantitatively, the slope parameter varies
by less than 0.1 \% in the range of particle masses $0 - 2$ GeV.
As a consequence, in this simple phase-space 
model the transverse mass spectra of all particle species have
the same slope. 

\begin{figure}[ht]
\begin{center}
\epsfig{file=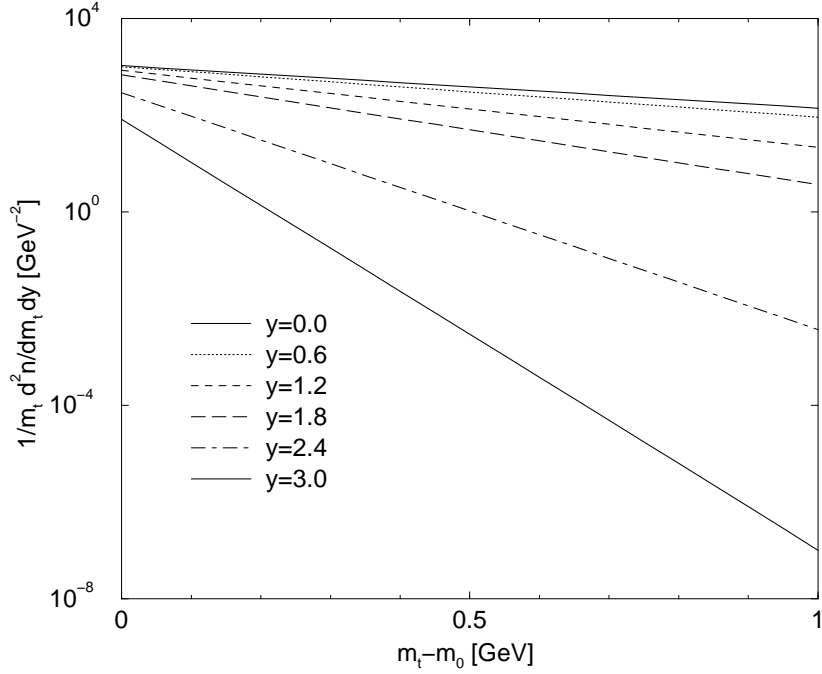,width=9cm,angle=-90}
\caption{\small Transverse mass spectra for $\pi^-$-mesons in
Pb+Pb-reactions at 158 AGeV.}
\label{fig5}
\end{center}
\end{figure}

In Fig.\ \ref{fig5} we demonstrate the exponential decay of the
transverse mass spectra by
showing $\pi^-$ transverse mass spectra for
different rapidities, computed from Eq.\ (\ref{eqdndmdy8}) with $a=0$ and
using the particle multiplicities of Table \ref{tab3} to compute the
phase-space integrals.
One observes that the slope parameter increases with increasing rapidity,
as is obvious from Eq.\ (\ref{eqT1}).

The case where $a=0$ and all particles are massless deserves a special
treatment.
If all particles have the same mass, then the mean multiplicity
of the $n$th particle is simply the total multiplicity,
$\langle n_n \rangle \equiv n$, and we obtain for the transverse
mass spectrum (omitting the index $n$) \cite{Risch}
\be
\frac{1}{m_{\perp }}\,\frac{d^2n}{dm_{\perp} dy}
= \pi \, n \, \frac{I_{n-1}(s')}{I_n(s)}
= \frac{2n(n-1)(n-2)}{s} 
\left(1-\frac{2\,m_\perp \cosh y}{\sqrt{s}}\right)^{n-3}  \;,
\ee[trmm=0exa]
where we have used Eqs.\ (\ref{exaresult}) and (\ref{eqs'}).
Using the saddle-point condition $\beta \equiv 2n/ \sqrt{s}$,
cf.\ Eq.\ (\ref{spm=0}), and $(1-x/n)^n \simeq e^{-x}$ (valid for
large $n$), we obtain
\be
\frac{1}{m_{\perp }}\,\frac{d^2n}{dm_{\perp} dy}
\simeq n\, \frac{\beta^2}{2} \, e^{- \beta \,m_\perp \cosh y}\;,
\ee[trmm=0]
where we have dropped subleading terms of order $O(1/n)$.
One observes that the transverse mass spectrum is exponentially
decreasing as a function of the particle energy 
$E \equiv m_\perp \cosh y$. From Eq.\ (\ref{spm=0}) we see that the
inverse slope parameter $1/\beta$ corresponds to half of the available energy
per particle. We note that the result (\ref{trmm=0}) could have also
be obtained from Eqs.\ (\ref{eqdndmdy8}) and (\ref{eqT1}). To see
this, employ $a=0$ and Eq.\ (\ref{exaresult}) to show that $1/T^* = \beta$.
Note that the apparent ``temperature'' $T^*$ is half
the energy per particle, $T^*= \frac{1}{2}\, \sqrt{s}/n$ \cite{Fri,Risch}.
This is in contrast to the temperature $T$
for an ultrarelativistic gas in thermal equilibrium, which appears
instead of $T^*$ in the transverse mass spectrum, see Eq.\ (\ref{eqEdNd3p})
below, and which is one third of the thermal energy per particle, 
$T=\frac{1}{3}\, E/n$. Thus, if we compare transverse mass
spectra for the phase-space model and the thermal model
at the same total energy, $\sqrt{s} = E$, the
apparent temperature is a factor $3/2$ times the thermal 
temperature, $T^*  = \frac{3}{2} \, T$. The inverse slope parameter
$T^*$ of the transverse mass spectra should thus not be
interpreted as the temperature in thermodynamical equilibrium.

Integrating either the exact result (\ref{trmm=0exa}) or the approximate
result (\ref{trmm=0}) for the transverse mass spectrum of massless particles
over $m_\perp$, we obtain the rapidity spectrum
\be   
\frac{dn}{dy}= \frac{n}{2\cosh^2y}  \;.
\ee[eqdndy1] 
The half width $\Delta y$ of the rapidity spectrum for massless particles is 
independent of the particle number and the 
total energy $\sqrt{s}$ of the reaction, $\Delta y = \ln ( 1 +
\sqrt{2}) \simeq 0.8814$. 
A further integration gives
\be  
\int dy\, \frac{dn}{dy}= n \int \frac{dy}{2\cosh^2y}=n\;,  
\ee[13]
as expected.

\subsection{The thermal model} 

We shall compare our transverse mass spectrum computed from the
phase-space integral with those obtained in a thermal model.
The latter are obtained from the Cooper-Frye formula \cite{Coop}
\be   
E_n\,\frac{dn_n}{d^3p_n}
\equiv \frac{1}{m_{\perp n}}\,
\frac{d n_n}{dm_{\perp n} dy_n} 
=\int_\Sigma d\sigma_\nu p^\nu_n \, g_n(x,p_n)\;,
\ee[eqd3Nd3p]
which counts the number of on-shell particles with three-momentum ${\bf p}_n$
streaming through a space-time hypersurface $\Sigma$ with normal
vector $d \sigma_\mu$. Here, $g_n(x,p_n)$ is the single-particle
on-shell momentum distribution function for the $n$th particle at 
space-time point $x$ and
$\Sigma$ is the hypersurface where kinetic freeze-out of particles occurs.
In general, this hypersurface could have a complicated shape in space-time.
Here, we make the simplifying assumption that freeze-out occurs at
fixed time in the center-of-momentum frame, 
such that $d\sigma_\nu=(d^3x,\mathbf 0)$. Provided
that $g_n(x,p_n)$ does not change over the freeze-out hypersurface,
the integration
over $d \sigma_\nu$ then simply yields a factor $V$, the total
volume of the system. We also take
the particles to be distributed according to the appropriate 
quantum statistical distribution function in thermal equilibrium, 
$g_n(x,p_n) = d_n/(2 \pi)^3 \, (e^{(E_n-\mu_n)/T} \pm 1)^{-1}$, 
where $d_n$ is the degeneracy and $\mu_n$ the chemical potential.
The upper sign holds for fermions and the lower for bosons. The
invariant momentum spectrum (\ref{eqd3Nd3p}) reduces to
\be 
\frac{1}{m_{\perp n}}\,
\frac{dn_n}{d m_{\perp n} dy_n}= \frac{d_n \, V}{(2 \pi)^3}\, 
\frac{m_{\perp n} \cosh y_n}{
e^{(m_{\perp n} \cosh y_n-\mu_n)/T} \pm 1}\;.  
\ee[eqEdNd3pQS]
For particle energies $E_n =m_{\perp n} \cosh y_n \gg T, \, \mu_n$, 
this becomes the classical Boltzmann distribution,
\be 
\frac{1}{m_{\perp n}}\,
\frac{dn_n}{d m_{\perp n} dy_n} \simeq \frac{d_n V e^{\mu_n/T}}{(2 \pi)^3}\, 
m_{\perp n}\, \cosh y_n
\, e^{-m_{\perp n} \cosh y_n/T}\;.  
\ee[eqEdNd3p]
Note that this thermal distribution is not an exponential in
the particle energy
$E_n= m_{\perp n} \cosh y_n$ 
because of the prefactor $m_{\perp n} \cosh y_n$ which introduces logarithmic
corrections to the pure exponential decay.

In this work, we do not attempt to fit the temperature, the volume, 
and the chemical potential to reproduce the particle multiplicities.
There is a vast amount of literature regarding such fits, see e.g.\ 
\cite{Braun,Becca,Cley}. Our goal is to compare the invariant phase space 
available to the system at a given energy to the corresponding
thermodynamic phase space. 
To this end, we note that all hadrons except pions are sufficiently heavy
so that the Boltzmann approximation (\ref{eqEdNd3p})
for the particle spectra is applicable. In this case, the
chemical potential only enters the overall normalization of the
invariant momentum spectrum, but does not change its shape.
For pions, on the other hand, the chemical potential is in general
quite small \cite{Braun,Becca,Cley}. It is therefore justified to use the
following ansatz for the particle spectra
\be
\frac{1}{m_{\perp n}}\,
\frac{dn_n}{d m_{\perp n} dy_n} = c_n\, 
\frac{m_{\perp n} \cosh y_n}{
e^{m_{\perp n} \cosh y_n/T} \pm 1}\;,
\ee[trmapp]
where $c_n$ contains the information about the volume of the system,
the degeneracy, and the chemical potential and
is determined from the multiplicity of the $n$th particle,
\be   
\langle n_n \rangle= c_n\, \int d^3p_n \, \frac{1}{e^{E_n/T} \pm 1}\;.
\ee[eqN]
The rapidity spectrum $dn_n/dy_n$ is obtained from Eq.\ (\ref{trmapp})
by integrating over $m_{\perp n}$.

In order to compute the invariant momentum spectrum (\ref{trmapp}),
we still need to fix the temperature.
We choose it such that the total thermal energy $E$ is
equal to the center-of-momentum energy $\sqrt{s}$ available
for particle production,
\be  
\sqrt{s} \equiv E = \sum_{r} \int d^3 p_r \, E_r \,
\frac{1}{m_{\perp r}}\,
\frac{dn_r}{d m_{\perp r} dy_r} =
\sum_{r} c_r \int d^3 p_r\,
\frac{E_r}{\exp(E_r/T)\pm 1} \;,
\ee[eqE2]
where $r$ labels the particle species.

\section{Comparison with Experimental Data} \label{IV}

In this section, we compute transverse mass and rapidity spectra
from both the phase space as well as the thermal model and
compare them to the experimental data. 
 
\subsection{Completing particle multiplicities in central Pb+Pb
collisions at 158 AGeV} 

The calculation of
the phase-space integral requires knowledge of the multiplicities
of the various particle species, their masses, 
and the total energy of the reaction.
Since the NA49 experiment does not measure all particle species produced
in a nuclear collision, 
the multiplicity data \cite{NA49,Afan,Misch,NA49lam,NA49xi,NA49ome} must be 
completed by using symmetry considerations, electric charge, baryon
number and strangeness conservation, as well as empirical rules.

We assume that the $n$ particles in the final state are solely produced
by strong interactions. Therefore, the multiplicities measured in
the experiment have to be corrected for weak decays. This has already
been done in the data of the NA49 collaboration \cite{NA49}. We show the
measured multiplicities for pions, kaons, $\Lambda$'s, $\Xi$'s, and
$\Omega$'s in Table \ref{tab2}.

\begin{center}
\begin{table}[h]
\begin{tabular}{|l|r|r|cl|r|r|}\hline
Hadron & $m_r$[GeV] & $\langle n_r \rangle $ & \hspace*{0.2cm}
& Hadron & $m_r$[GeV] & $ \langle n_r \rangle$  \\ \hline
$\pi^+$ & 0.1396 & 619.0 & & $\bar{\Lambda}$ & 1.1157 & 4.64  \\
$\pi^-$ & 0.1396 & 639.0 & &$\Xi^-$ & 1.3213 & 4.45 \\
$K^+$ & 0.4937 & 103.0   & &$\Xi^+$  & 1.3213 & 0.83 \\
$K^-$ & 0.4937 & 51.9  &  & $\Omega$ & 1.6725 & 0.62 \\
$K^0_S$ & 0.4977 & 81.0 & & $\bar{\Omega}$ & 1.6725 & 0.20 \\  
$\Lambda$ & 1.1157 & 53.0 & &  & & \\ \hline
\end{tabular}
\caption{Multiplicities of different particle species measured 
by the NA49 collaboration \cite{NA49} in Pb+Pb collisions at 158 AGeV and
corrected for weak decays.}
\label{tab2}
\end{table}
\end{center}

Based on Table \ref{tab2}, we now reconstruct
the multiplicities for particle species not directly measured by NA49.
\begin{itemize}
\item $\pi^0$: 
Due to (approximate) isospin symmetry in the reaction, we
take 
\be 
\langle \pi^0 \rangle = \frac{\langle \pi^+ \rangle
+ \langle \pi^- \rangle }{2}\;.
\ee[14]
\item $K^0,\bar{K}^0$: 
We assume that $\langle K_S^0 \rangle= \langle K_L^0 \rangle$ and
$\langle K^0 \rangle= \langle \bar{K}^0 \rangle$.
Since $\langle K^0 \rangle + \langle \bar{K}^0 \rangle
= \langle K_S^0 \rangle + \langle K_L^0 \rangle$, we have
\be
\langle K^0 \rangle= \langle \bar{K}^0 \rangle
= \langle K_S^0 \rangle\;.
\ee[15]
\item $\Lambda,\Sigma^+,\Sigma^0,\Sigma^-:$ 
An empirical rule found by Wroblewski \cite{Wrob,Gazdz} states that
\be  
\langle \Sigma^+ \rangle + \langle \Sigma^- \rangle
=0.6 \left[ \langle \Lambda \rangle + \langle \Sigma^0 \rangle \right]\;.
\ee[sigma+-]
Here, $ \langle \Lambda \rangle + \langle \Sigma^0 \rangle$ 
corresponds to the number of {\em measured\/} $\Lambda$'s as given
in Table \ref{tab2}. Note that the measured $\Lambda$ multiplicity
is not the {\em true\/} multiplicity, $\langle \Lambda \rangle$,
because the $\Sigma^0$ decays to $100\%$ into a $\Lambda$. One therefore has
to reduce the measured $\Lambda$ multiplicity by that of 
the $\Sigma$'s.
Equation (\ref{sigma+-}) yields the combined multiplicity of $\Sigma^+$ and
$\Sigma^-$ directly from the measured data.
Due to isospin symmetry, we may assume that
\be 
\langle \Sigma^0 \rangle=\frac{\langle \Sigma^+ \rangle+
\langle\Sigma^- \rangle}{2}   \;.
\ee[sigma0]
Now the multiplicities of $\Sigma^+ + \Sigma^-$ and $\Sigma^0$
are known. 
Combining Eqs.\ (\ref{sigma+-}) and (\ref{sigma0}), we
can solve for $\langle \Lambda \rangle$,
\be
\langle \Lambda \rangle = \frac{7}{3}\, \langle \Sigma^0
\rangle = \frac{7}{6} \left[
\langle \Sigma^+ \rangle + \langle \Sigma^- \rangle \right] =
0.7 \left[ \langle \Lambda \rangle + \langle \Sigma^0 \rangle \right]\;.
\ee[16]
The multiplicities for the corresponding antiparticles are
obtained in the same way.

\item $\Xi^0, \bar{\Xi}^0$: Due to isospin symmetry we have
\be  
\langle \Xi^0 \rangle = \langle \Xi^- \rangle \;, \;\;\;\;
\langle \bar{\Xi}^0 \rangle = \langle \Xi^+ \rangle  \;.
\ee[17]

\item Nucleons:
The net number of protons and neutrons follows from baryon number
conservation,
\bea  
\lefteqn{\langle p \rangle - \langle \bar{p} \rangle
+ \langle n \rangle - \langle \bar{n} \rangle} \nonumber \\
& = & N_w - \left[ \langle \Lambda \rangle - \langle \bar{\Lambda}
\rangle + \langle \Sigma^+ + \Sigma^- \rangle - \langle \bar{\Sigma}^+
+ \bar{\Sigma}^- \rangle + \langle \Xi^0 \rangle
- \langle \bar{\Xi}^0 \rangle +
\langle \Xi^- \rangle - \langle \Xi^+ \rangle 
+ \langle \Omega^- \rangle - \langle \Omega^+ \rangle \right]
\nonumber \\
& = & 276.96\;,
\eea[18]
where we have used the number of wounded nucleons $N_w$ and
the previously determined multiplicities.
Assuming isospin symmetry, 
\be
\langle p \rangle =\langle n\rangle\;, \;\;\;\;
\langle \bar{p} \rangle = \langle \bar{n} \rangle\;,
\ee[19]
we get $\langle p \rangle - \langle \bar{p} \rangle=
\langle n \rangle - \langle \bar{n} \rangle = 138.48$.
In order to obtain the individual multiplicities, we assume
that the production of strange antiquarks relative to strange quarks
is suppressed as compared
to the production of nonstrange antiquarks relative to nonstrange
quarks
by a factor $D\equiv \frac{\langle \bar{s} \rangle  \langle q \rangle}{
\langle s \rangle \langle \bar{q} \rangle}$ \cite{Raf,Bial1}. This
results in
\be
\frac{\langle \bar{p} \rangle}{\langle p \rangle}
= D^{-1}\,\frac{\langle \bar{\Lambda} \rangle}{\langle \Lambda \rangle}
=D^{-2}\, \frac{\langle \Xi^+ \rangle}{\langle \Xi^- \rangle}
= D^{-3}\,\frac{\langle \Omega^+ \rangle}{\langle \Omega^- \rangle}\;.
\ee[21]
$D$ can be obtained from the measured multiplicities.
We take the average value of
$\frac{\langle \Xi^+ \rangle}{\langle \Xi^- \rangle}/ 
\frac{\langle \bar{\Lambda} \rangle}{\langle \Lambda \rangle}$ 
and $\frac{\langle \Omega^+ \rangle}{\langle \Omega^- \rangle}/
\frac{\langle \Xi^+ \rangle}{\langle \Xi^- \rangle}$, respectively.
This yields $D= 1.93$.
The antiproton-to-proton ratio is then 
$\frac{\langle \bar{p} \rangle}{\langle p \rangle}=0.0454$.

\end{itemize}
The completed multiplicities are summarized in Table \ref{tab3}.
The average number of the secondaries thus obtained is 2611.
Note that we do not attempt to reconstruct resonances decaying
strongly into pions and nucleons, such as the $\rho$ and the $\Delta$.
In order to determine the corresponding multiplicities one would
require additional assumptions, e.g.\ thermodynamical equilibrium
between various particle species. As we aim to present an approach
that is complementary to the thermal model, we refrain from doing so.
Moreover, one would then have to treat the
decay of these reconstructed resonances separately in order to compare with the
measured spectra which already contain decay products.
We leave this for a future improvement of our approach and,
in this work, do not make a distinction between
direct particle production and subsequent resonance decay.

\begin{center}
\begin{table}
\begin{tabular}{|l|r|r|cl|r|r|} \hline
Hadron & $m_r$[GeV] & $\langle n_r \rangle$ & \hspace*{0.2cm}
&Hadron & $m_r$[GeV] & $\langle n_r \rangle$  \\ \hline
$\pi^+$ & 0.1396 & 619 && $\bar{\Sigma}^+ +\bar{\Sigma}^-$ & 1.1934 & 3    \\
$\pi^-$ & 0.1396 & 639 && $\Xi^-$ & 1.3213 & 4 \\
$\pi^0$ & 0.1350 & 629 && $\Xi^+$ & 1.3213 & 1 \\
$K^+$ & 0.4937 & 103   && $\Xi^0$ & 1.3148 & 4 \\
$K^-$ & 0.4937 & 52    && $\bar{\Xi}^0$ & 1.3148 & 1 \\
$K^0+\bar{K}^0$ & 0.4977 & 162 &&  $\Omega + \bar{\Omega}$ & 1.6725 & 1 \\  
$\Lambda$ & 1.1157 & 37 &&  $p$ & 0.9383 & 145 \\
$\bar{\Lambda}$ & 1.1157 & 3 && $\bar{p}$ & 0.9383 & 7 \\
$\Sigma^+ +\Sigma^-$ & 1.1934 & 32 && $n$ & 0.9396 & 145 \\
$\Sigma^0$ & 1.1926 & 16  && $\bar{n}$ & 0.9396 & 7 \\
$\bar{\Sigma}^0$ & 1.1926 & 1 & &  &  & \\ \hline
\end{tabular}
\caption{Completed list of multiplicities.}
\label{tab3}
\end{table}
\end{center}

\subsection{Transverse mass and rapidity spectra from 
phase space, {\boldmath $a=0$}}

In this section, we compute pion and kaon spectra from
the phase-space model with $a=0$ and compare them to experimental
NA49 data from the $5\%$ most central Pb+Pb collisions at 
158 AGeV.

\begin{figure}[ht]
\begin{center}
\epsfig{file=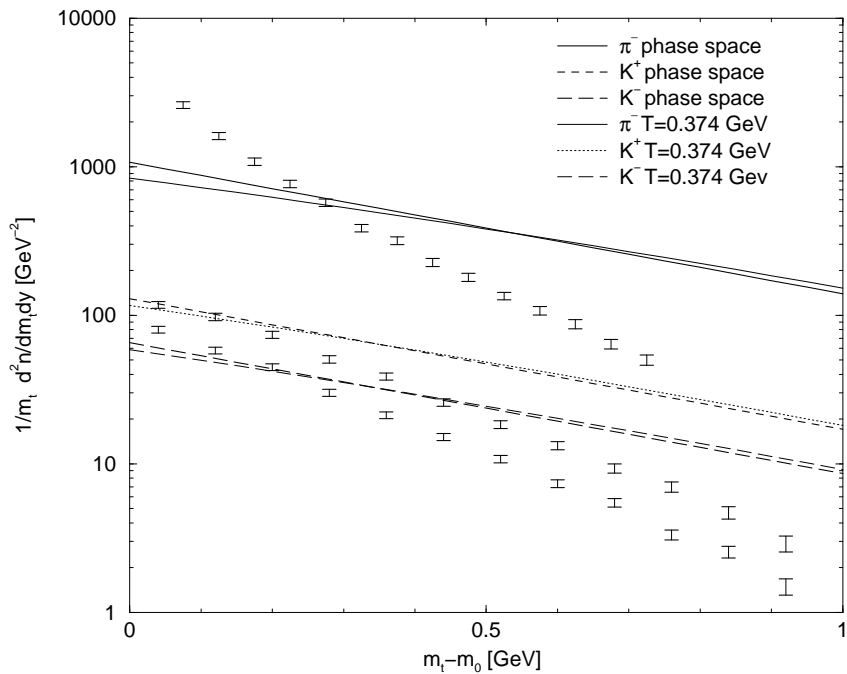,width=9cm,angle=-90}
\caption{\small Transverse mass spectra for $\pi^-, K^+,$ and 
$K^-$ calculated from the phase-space model with $a=0$, as
well as from a thermal model.
The pion data is taken in the range $0 < y < 0.2$, 
and the kaon data for $|y| < 0.1$. 
Experimental data is shown by the symbols.
Here, as in the following figures, 
error bars include statistical errors (as given by NA49)
and a systematical error of $5\%$.}
\label{fig8}
\end{center}
\end{figure}

Figure \ref{fig8} shows transverse mass 
spectra for $\pi^-$, $K^+$, and $K^-$ mesons,
calculated from the phase-space model as well
as from a thermal model, together with the measured data. 
We observe that the phase-space model and the thermal 
model give quite similar results. The agreement becomes
better for larger particle mass. This is not surprising, since
the nonrelativistic limit of relativistic invariant phase space,
$\int d^3{\bf p}_i/ E_i \rightarrow (1/m_i) \int d^3{\bf p}_i$,
is identical to the phase space in the thermal model.

However, both models disagree with the data. The slope is too small, 
i.e., the inverse slope parameter is too large.
In the phase-space model, 
$T^*_\pi=0.4935$ GeV and $T^*_K=0.4934$ GeV. In the thermal
model, the temperature parameter extracted from the
energy constraint (\ref{eqE2}) is $T=0.374$ GeV for both
pions and kaons. Note that the inverse slope parameter
$T^*$ is larger than the temperature $T$, $T^*/T \simeq 1.32$. 
In the ultrarelativistic limit, $T^*/ T = \frac{3}{2}$,
cf.\ discussion above. In the nonrelativistic limit, i.e., for
$m_i \gg T$, $T^*/ T \rightarrow 1$, since then invariant phase space
becomes identical to thermodynamic phase space. 
Pions and kaons are neither ultrarelativistic nor nonrelativistic,
which explains the value $T^*/T \simeq 1.32$, i.e., between 1 and
$3/2$. In the range of $m_{\perp}$ values shown in Fig.\ \ref{fig8},
both spectra appear to have a similar slope, but at larger
$m_\perp$, the thermal model spectrum decays faster than the
one calculated from the phase-space model, as it should, since
its inverse slope parameter $T$ is smaller than $T^*$ by $\sim 30 \%$.

\begin{figure}[ht]
\begin{center}
\epsfig{file=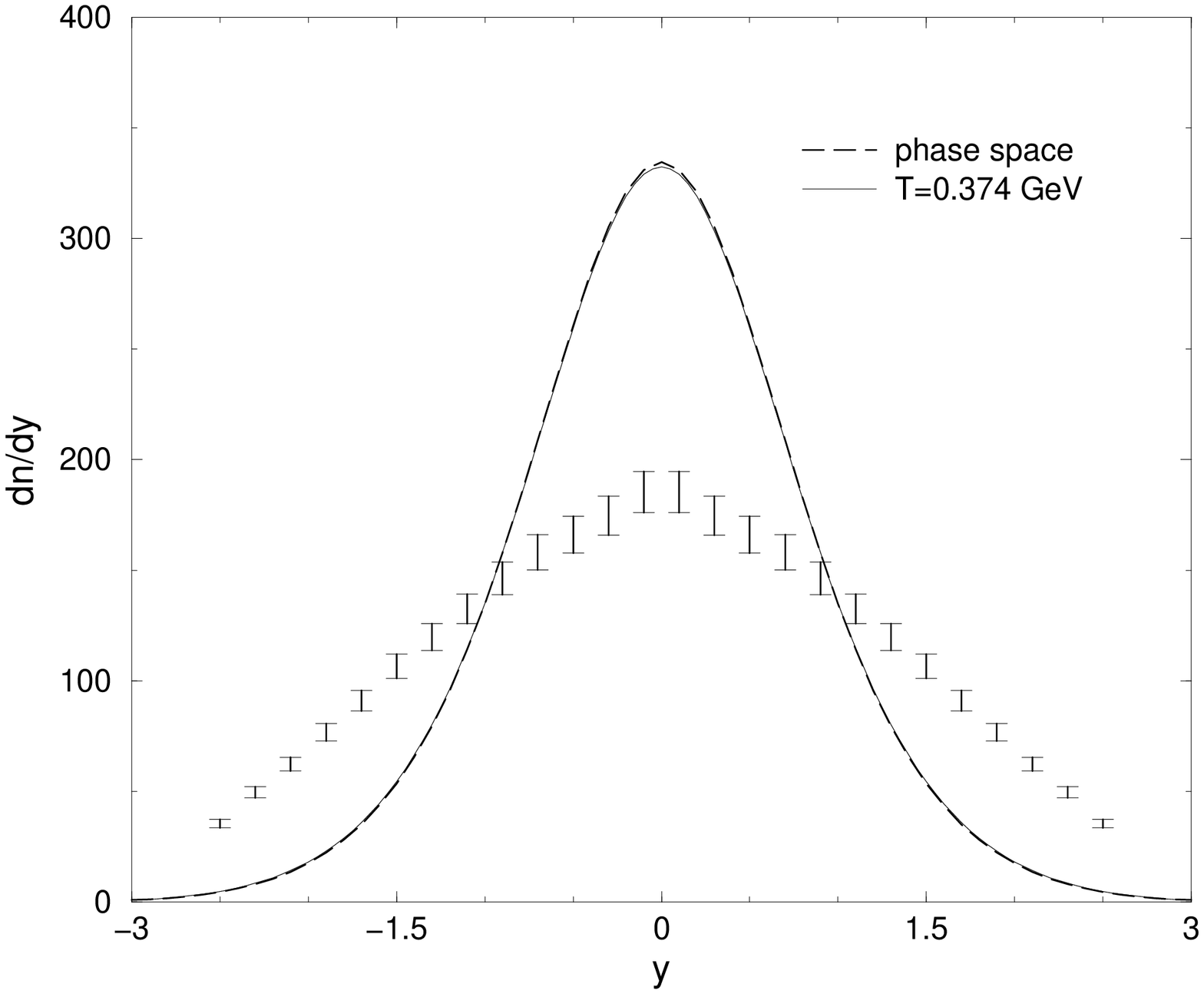,width=9cm,angle=-90}
\caption{\small Rapidity spectra for $\pi^-$ mesons, 
experimental data are shown by symbols, results
from the phase-space model by the dashed line and
from the thermal model by the solid line.}
\label{fig11}
\end{center}
\end{figure}

\begin{figure}
\begin{center}
\epsfig{file=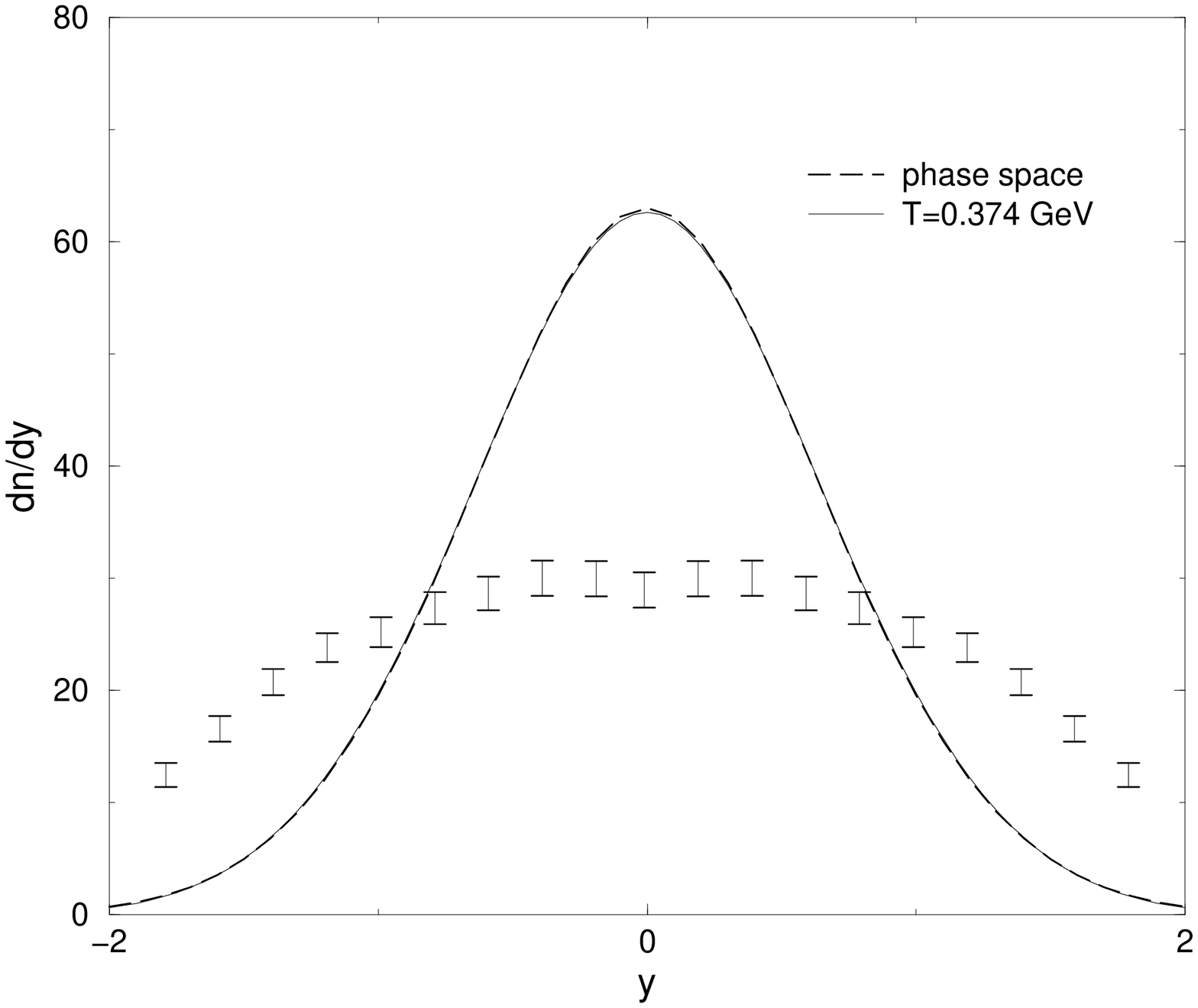,width=9cm,angle=-90}
\caption{\small The same as in Fig.\ \ref{fig11}
for $K^+$ mesons.}
\label{fig12}
\end{center}
\end{figure}

Let us now consider the rapidity spectra. 
Figure \ref{fig11} shows the measured spectrum of $\pi^-$ mesons 
in comparison to the phase-space model and the thermal
model. Phase-space and thermal model give
almost identical results. However, the calculated rapidity
spectra are much narrower than the measured one.
Similar results hold for $K^+$ mesons, see Fig.\ \ref{fig12}.

\begin{figure}[ht]
\begin{center}
\epsfig{file=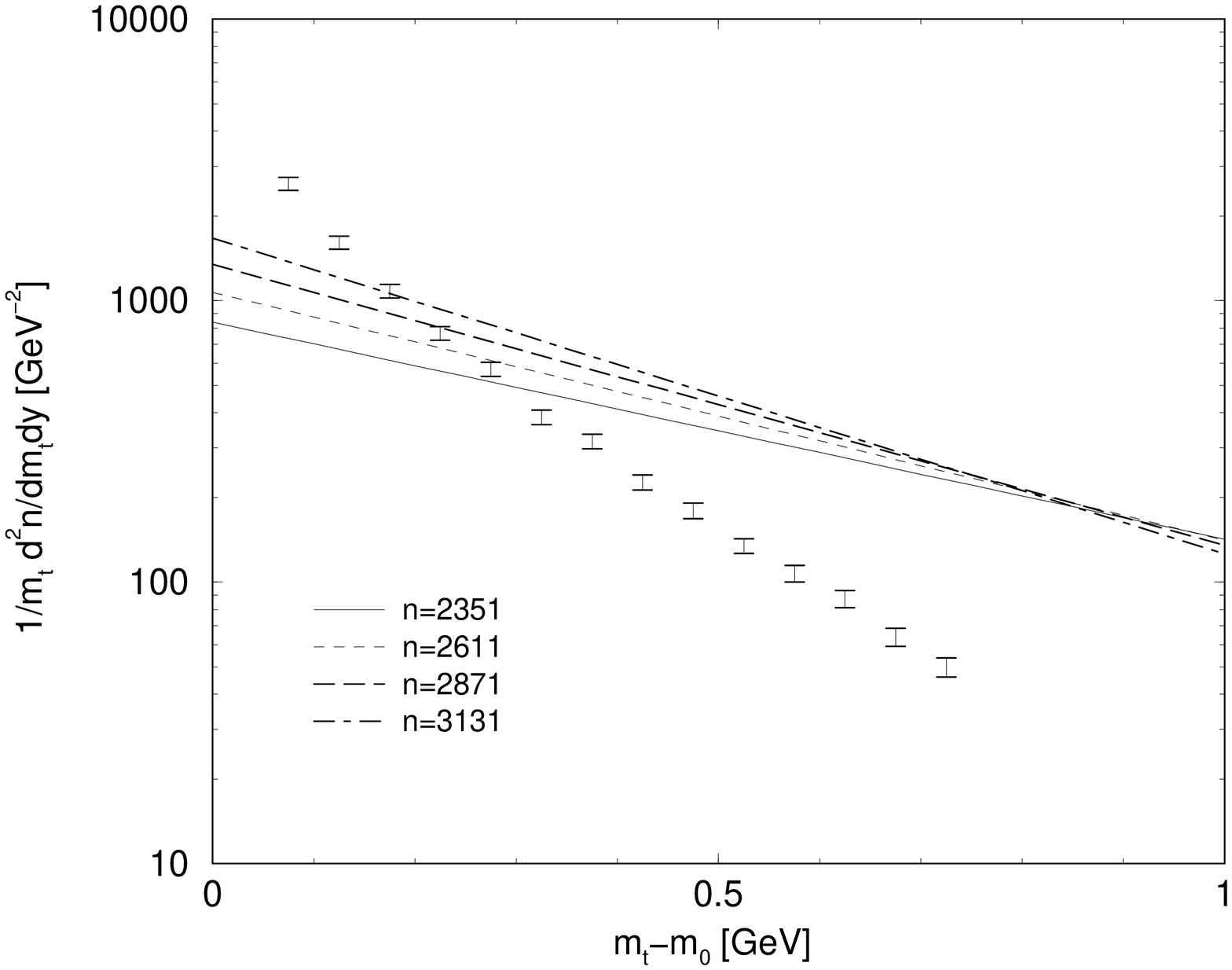,width=9cm,angle=-90}
\caption{\small Transverse mass spectrum for $\pi^-$ mesons for
different total multiplicities $n$.}
\label{fig19}
\end{center}
\end{figure}

\begin{figure}
\begin{center}
\epsfig{file=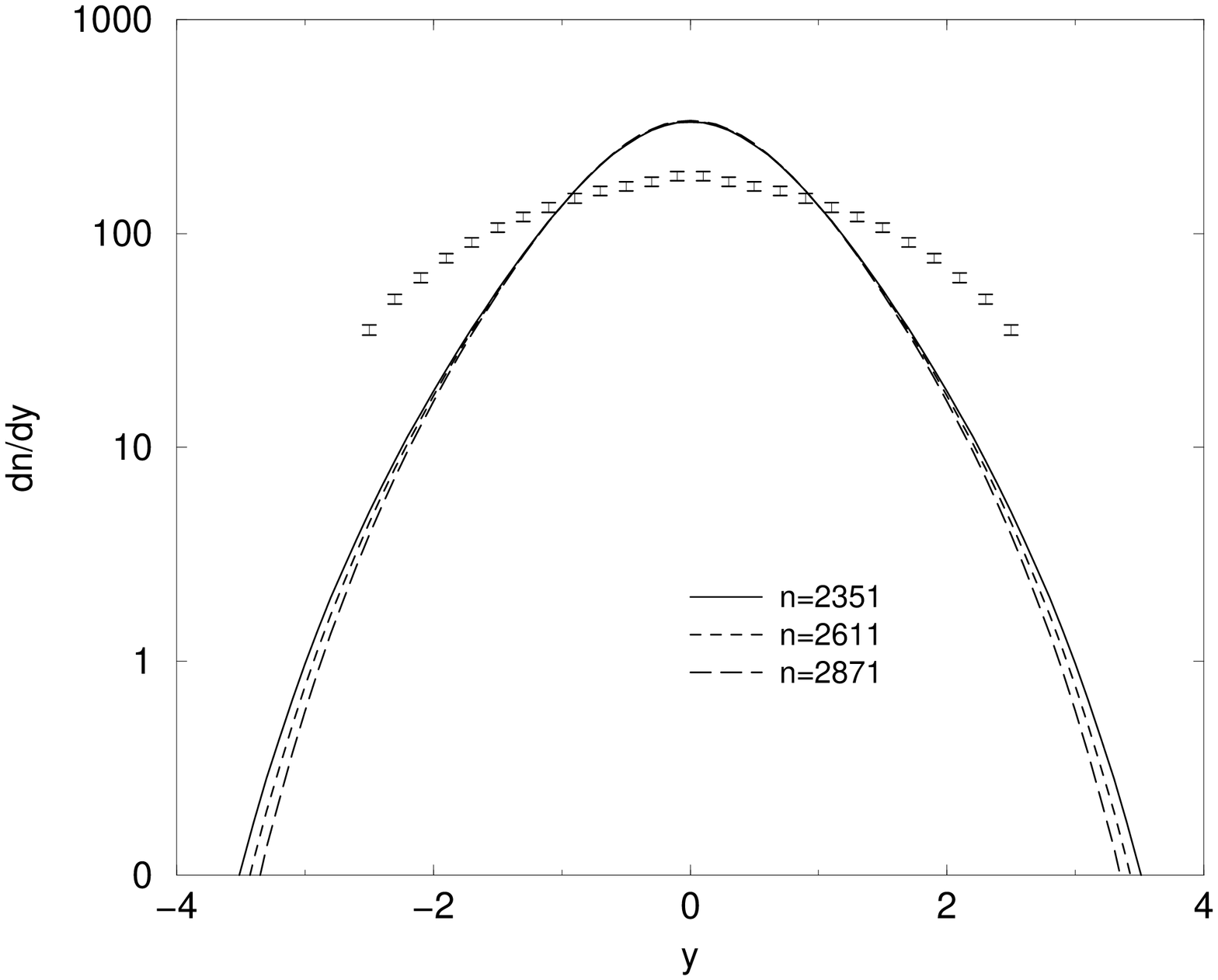,width=9cm,angle=-90}
\caption{\small Rapidity spectrum for $\pi^-$ mesons for
different total multiplicities $n$.}
\label{fig20}
\end{center}
\end{figure}

How sensitive are the calculated spectra to variations in the total
multiplicity? To answer this question, we performed two tests.
First, we added 200 massless particles and, alternatively, 
200 particles with mass 1 GeV, to the set of 2611 particles. 
For a given, sufficiently large $\sqrt{s}$, this
increases the derivative of $\ln I_{n-1}$ with respect to $\sqrt{s}$.
Then, according to Eq.\ (\ref{eqT1}),
the slope parameter $1/T^*$ of the transverse mass spectra increases.
This increase is independent of rapidity and the mass of the
considered particle species. It amounts to $7.5\%$ when adding massless
particles and to $14\%$ when adding particles with mass 1 GeV.
The difference between adding massless and massive particles can
be explained from Fig.\ \ref{fig4}: the increase of $\ln I_{n-1}$
with $\sqrt{s}$ is smaller for massless particles, thus, according to Eq.\
(\ref{eqT1}), the increase of
the slope parameter $1/T^*$ when adding particles is also smaller.
An uncertainty of 200 in the total multiplicity can therefore not
explain the deviation between the measured and the 
calculated transverse mass spectra for pions and kaons.

Second, we changed the total multiplicity by $k \times 260$ particles,
where $k= -1, 1, 2$, preserving the ratio between
the multiplicities of different particle species.
Figure \ref{fig19} shows the transverse mass spectra.
The inverse slope decreases with
increasing multiplicity, but even 560 additional particles
are not sufficient to reproduce the measured spectra.
In Fig.\ \ref{fig20} we show the corresponding rapidity spectra 
for $k=-1,0,1$ together with the experimental data in a 
logarithmic plot. 
The distributions become narrower with increasing multiplicity.
Thus, while the slope values of the transverse mass spectra approach the
data for increasing multiplicity, the rapidity distributions 
deviate more strongly.
It is therefore not possible to simultaneously
reproduce transverse mass spectra and rapidity spectra simply
by varying the total multiplicity.

\subsection{Transverse mass and rapidity spectra from 
phase space, {\boldmath $a \neq 0$}} 

We have seen that the transverse mass spectra calculated from
the phase-space model with $a=0$ do not agree with the experimental data:
their inverse slope parameter is too large. 
Similarly, the calculated rapidity spectra are too narrow.
This indicates a substantial remnant of the initial longitudinal 
motion of target and projectile, which still dominates the phase
space. Or in other words, complete isotropization has not
occurred.

The introduction of the factors $f_i({\bf p}_{\perp i})$, cf.\
Eq.\ (\ref{eqf}), with $a >0$, allows to suppress the transverse 
in favor of the longitudinal phase space. 
The parameter $a$ regulates the strength of 
the suppression of transverse momenta and has to be extracted from the
data. There are in principle two possibilities:
(1) $a$ is adjusted to reproduce the transverse mass spectra, or
(2) $a$ is adjusted to reproduce the rapidity spectra.

\begin{figure}[ht]
\begin{center}
\epsfig{file=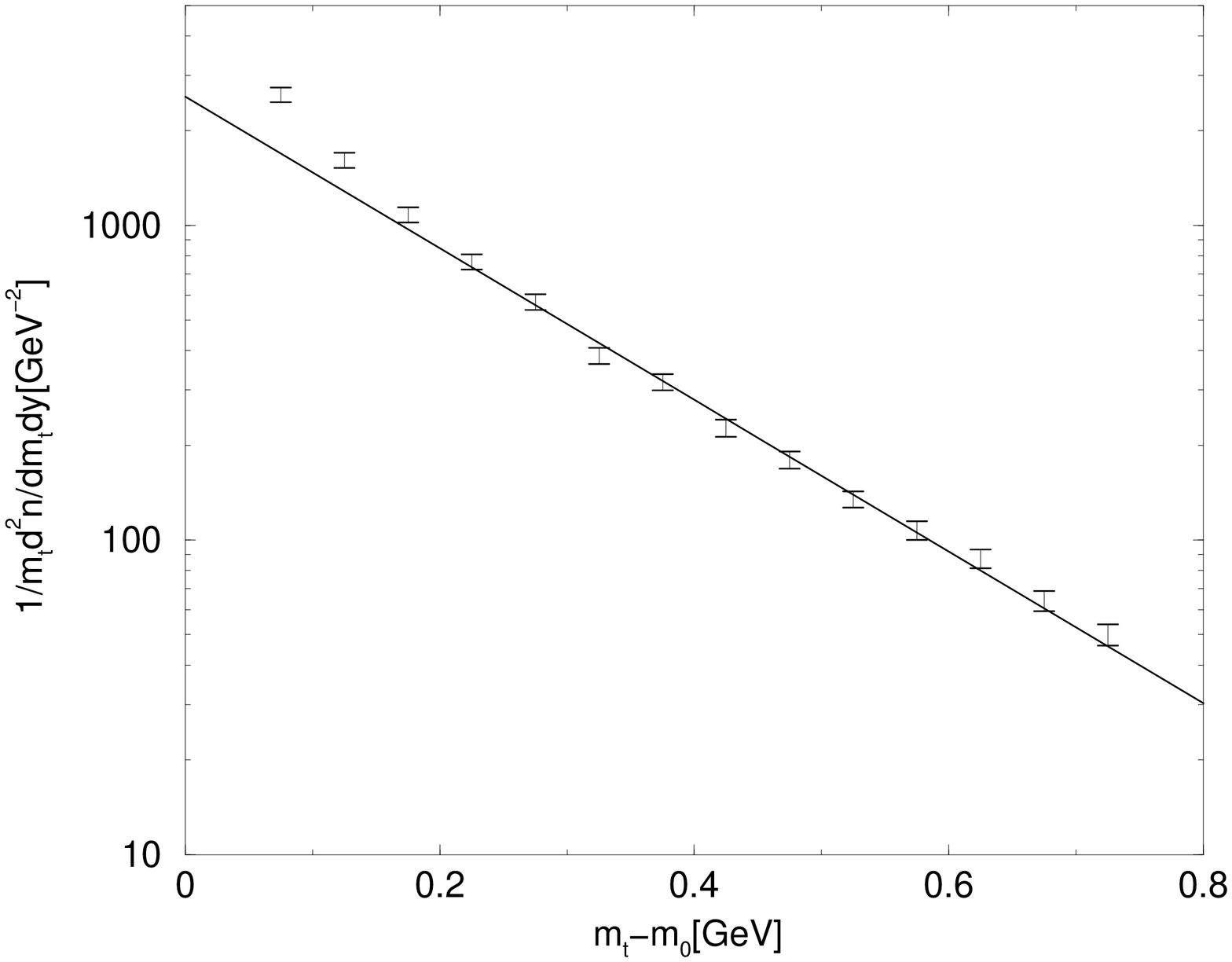,width=9cm,angle=-90}
\caption{\small Transverse mass spectrum for $\pi^-$ mesons in
the range $0 < y < 0.2$. The solid line shows the spectrum
calculated from the phase-space model for $a=(0.2\, \rm{GeV})^{-1}$.}
\label{fig22}
\end{center}
\end{figure}

\begin{figure}
\begin{center}
\epsfig{file=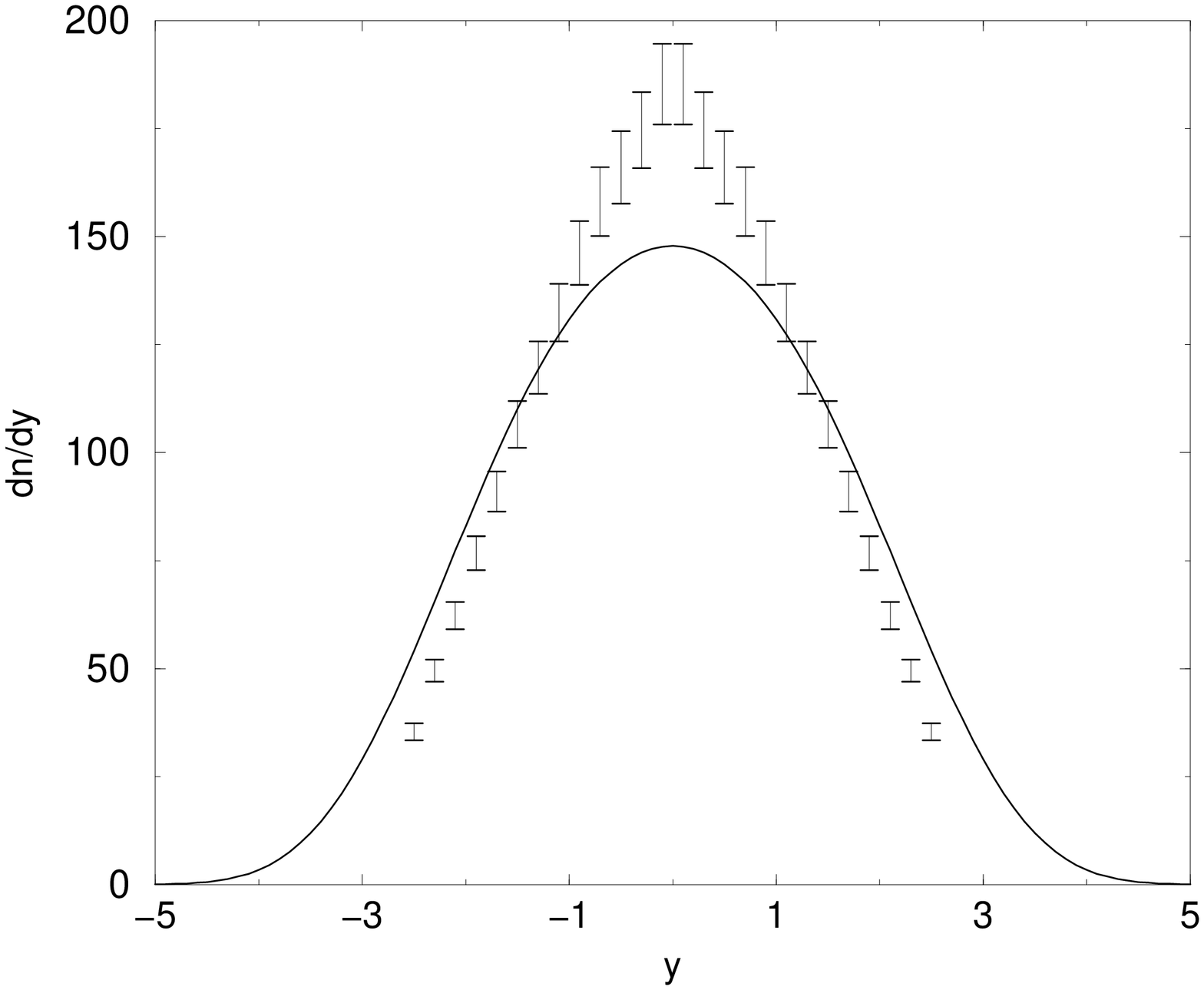,width=9cm,angle=-90}
\caption{\small The same as in Fig.\ \ref{fig22}, but
for the $\pi^-$ rapidity spectrum.}
\label{fig23}
\end{center}
\end{figure}

In Figs.\ \ref{fig22} and \ref{fig23} we show the transverse mass and
rapidity spectrum for $\pi^-$ mesons, calculated from the 
phase-space model in comparison with experimental data.
Choosing the value $a=(0.2\, {\rm  GeV})^{-1}$, we can quite well
reproduce the transverse mass spectrum. However, the
rapidity spectrum is then wider than the data, which indicates
that the suppression of transverse phase space is slightly too large.
Note that, at small transverse masses, 
the pion data lie above the calculated curve. This is quite naturally
explained by the fact that our simple phase-space 
model does not account for the distortion of phase space
due to resonance decays. Such resonance decays 
are expected to populate mainly the low-transverse
mass region where we observe the difference between the
data and the calculation \cite{knoll}.

\begin{figure}[ht]
\begin{center}
\epsfig{file=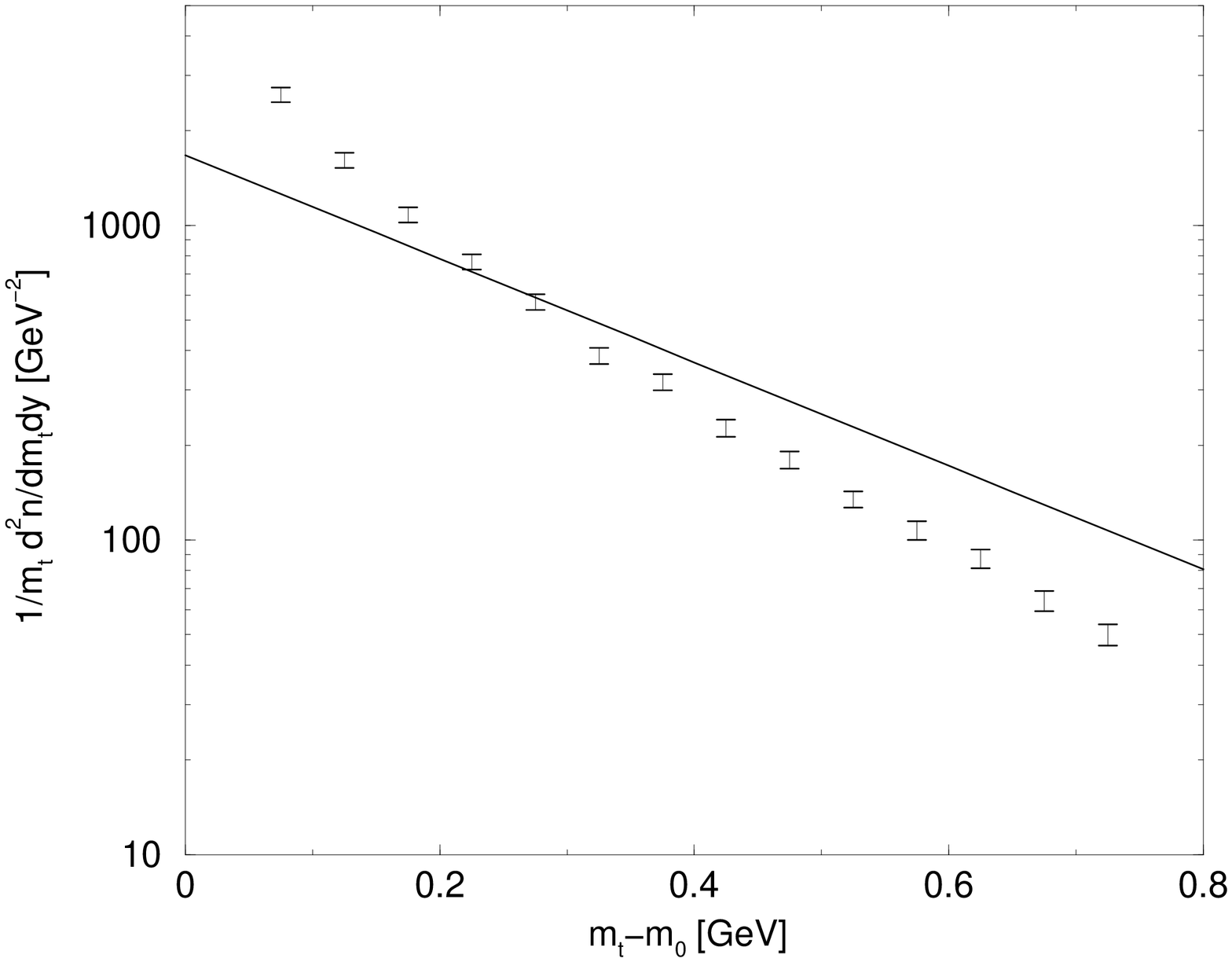,width=9cm,angle=-90}
\caption{\small The same as in Fig.\ \ref{fig22}, but 
with $a=(0.325\, \rm{GeV})^{-1}$.}
\label{fig28}
\end{center}
\end{figure}

\begin{figure}
\begin{center}
\epsfig{file=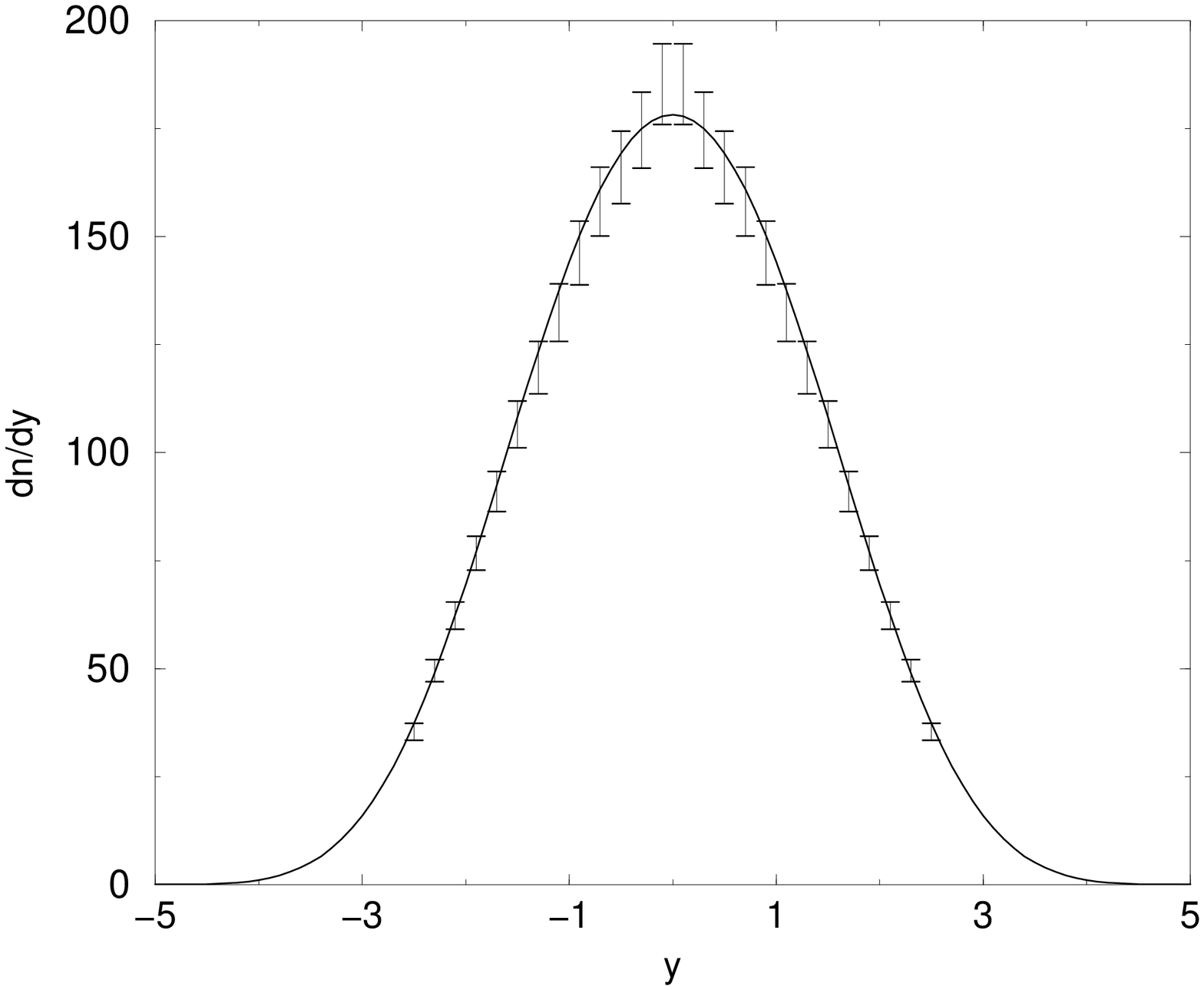,width=9cm,angle=-90}
\caption{\small The same as in Fig.\ \ref{fig23}, but
with $a=(0.325\, \rm{GeV})^{-1}$.}
\label{fig29}
\end{center}
\end{figure}

Figures \ref{fig28} and \ref{fig29} show the transverse mass and 
rapidity spectrum, now for a value $a=(0.325\, {\rm GeV})^{-1}$
which is adjusted to reproduce the rapidity spectrum. In this case,
the calculated transverse mass spectrum differs from the data
in that the inverse slope comes out too large. Transverse phase space
is not sufficiently suppressed in this case. 
Comparing the set of Figs.\ \ref{fig22}, \ref{fig23} with
the set of Figs.\ \ref{fig28}, \ref{fig29} suggests that there is a
value for $a$
between $(0.2 \, \rm{GeV})^{-1}$ and $(0.325\, \rm{GeV})^{-1}$, for
which a simultaneous fit of {\em both\/} transverse mass and 
rapidity spectrum, albeit of worse quality than the individual fits
shown in Fig.\ \ref{fig22} and Fig.\ \ref{fig29}, could be possible. 
Before returning to this question,
we summarize in Table \ref{tab4} the optimal fit values for
the parameter $a$ for $\pi^-$, $K^+$, and $K^-$ mesons.
In the first column, we quote the experimentally measured value
of the slope parameter of
the transverse mass spectrum, $T^*_{\rm exp}$.
In the second column we show the $a$-values optimized to
fit the transverse mass spectra, and in the last column those
optimized to fit the rapidity spectra. Note that, for the kaon
spectra, 
the two $a$ values thus obtained are quite close to each other. Indeed,
either a fit to the transverse mass spectrum or to the rapidity
spectrum will also yield a reasonable description of the respective
other spectrum.

\begin{center}
\begin{table}
\begin{tabular}{|c|c|c|c|} \hline
meson & $T^*_{\rm exp}$[GeV]   & $a^{-1} \, [{\rm GeV}^{-1}]$,
$m_\perp$-fit & 
$a^{-1} \,[{\rm GeV}^{-1}]$, $y$-fit   \\ \hline
$\pi^-$ & 0.180 & 0.200  &  0.325  \\
$K^+$ & 0.232 & 0.272 &  0.213 \\
$K^-$ & 0.226 & 0.263 &  0.315  \\   \hline
\end{tabular}
\caption{Optimal values for the damping parameter $a$. }
\label{tab4}
\end{table}
\end{center}

\begin{figure}[ht]
\begin{center}
\epsfig{file=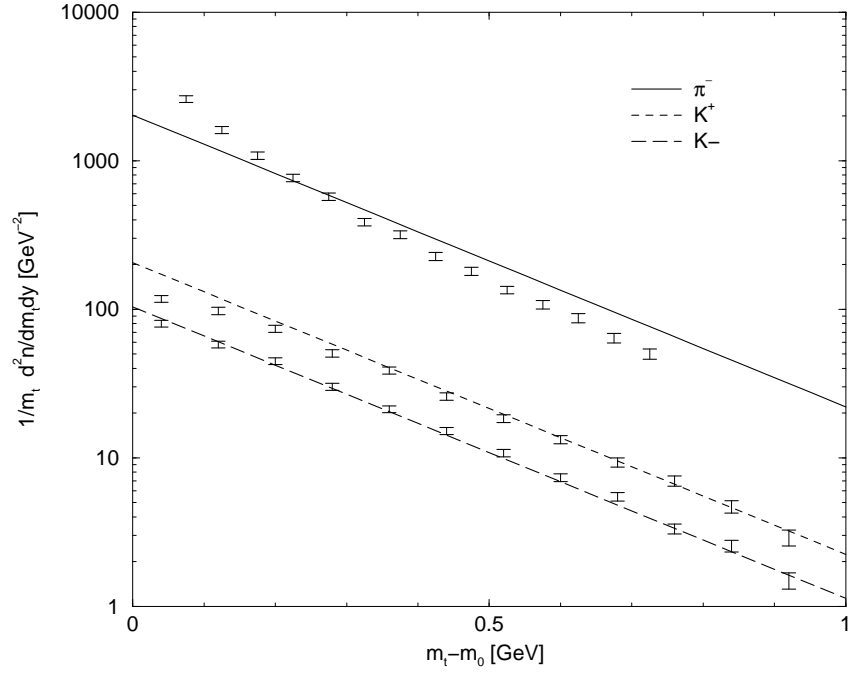,width=9cm,angle=-90}
\caption{\small Transverse mass spectra for $\pi^-$, $K^+$, and $K^-$ mesons.
The solid lines show the spectra calculated from the 
phase-space model with $a=(0.256 \, {\rm GeV})^{-1}$.} 
\label{fig39}
\end{center}
\end{figure}

\begin{figure}
\begin{center}
\epsfig{file=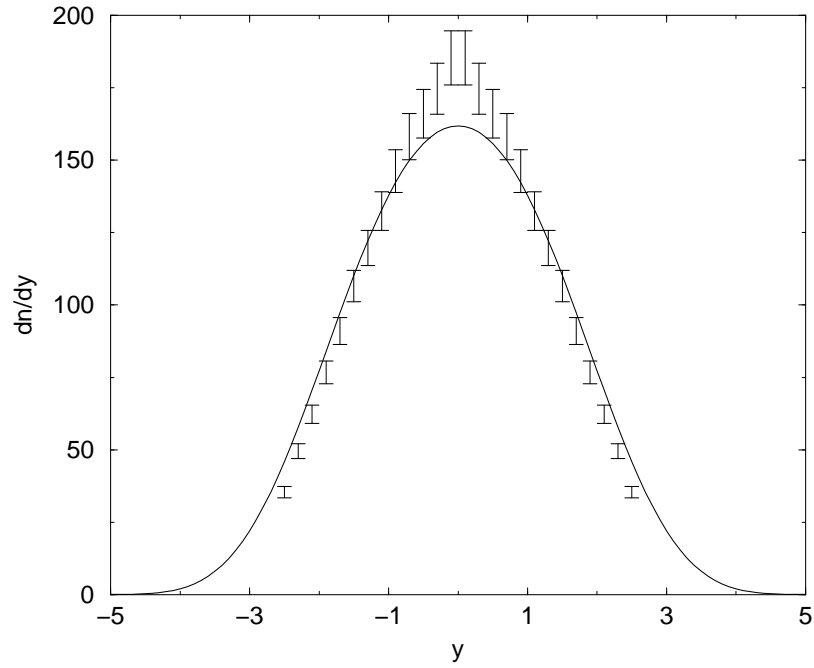,width=9cm,angle=-90}
\caption{\small Rapidity 
spectrum for $\pi^-$ mesons. The solid line shows the
spectrum calculated from the phase-space model with 
$a=(0.256\, {\rm GeV})^{-1}$.}
\label{fig40}
\end{center}
\end{figure}

\begin{figure}[ht]
\begin{center}
\epsfig{file=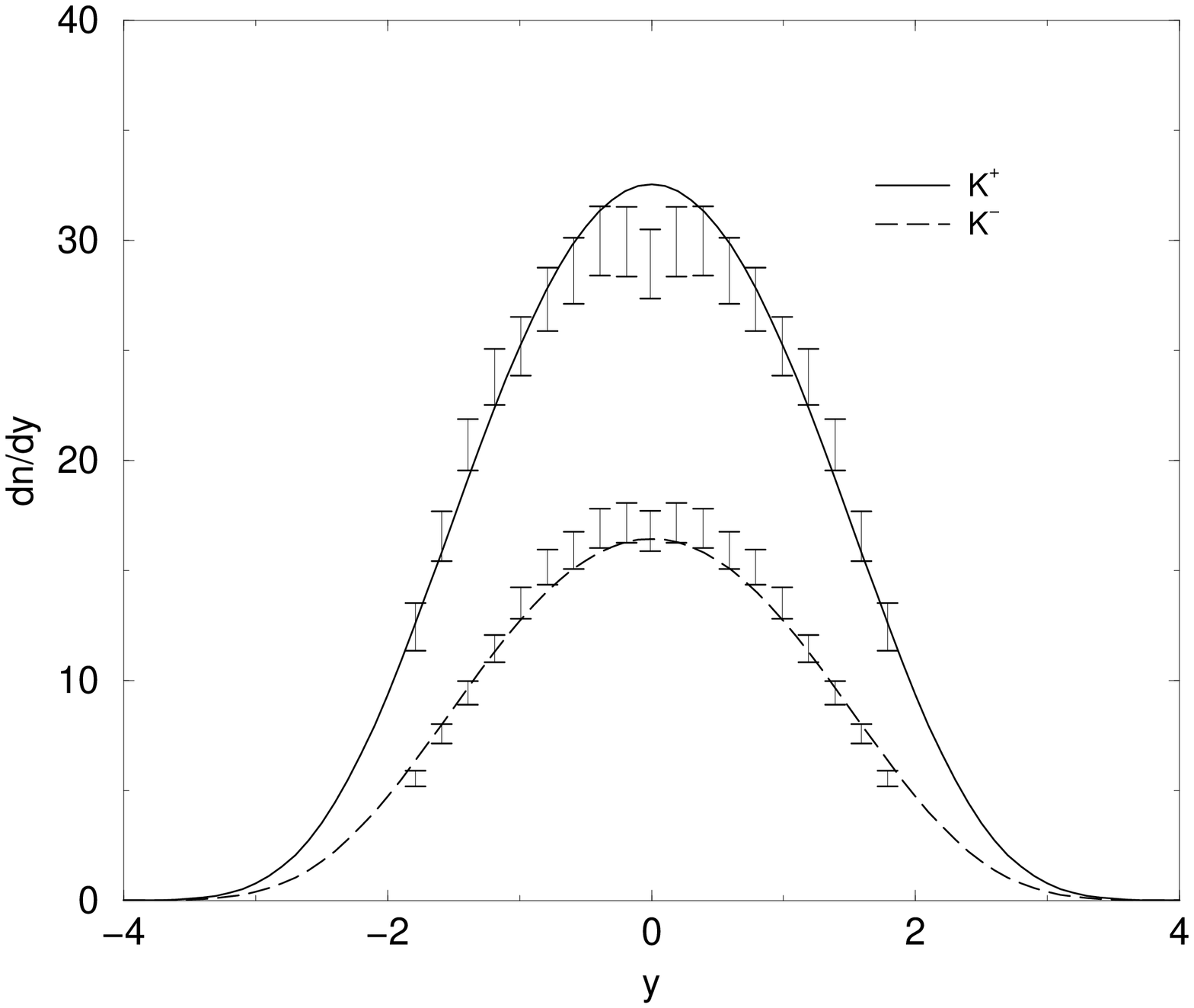,width=9cm,angle=-90}
\caption{\small The same as in Fig.\ \ref{fig40}, but for
kaons.}
\label{fig41}
\end{center}
\end{figure}

We now ask the question whether there is a single value for $a$ which
gives a reasonable description of both the transverse mass and 
the rapidity spectrum, and not only for pions but also for kaons. 
This is indeed the case, as shown in
Figs.\ \ref{fig39}, \ref{fig40}, and \ref{fig41}, for
$a=(0.256\, \rm{GeV})^{-1}$. 
It is quite remarkable that, for
a given set of particle multiplicities, a single parameter that
accounts for longitudinal phase-space dominance
is sufficient to yield a simultaneous fit of the transverse mass
and rapidity spectra for {\em both\/} pions {\em and\/} kaons. 
The observed deviations can be quite easily explained.
In the case of the pions, resonance (e.g.\ $\rho$ and $\Delta$) decays
populate the region of small transverse momenta and thus lead to
an enhancement of the measured as compared to the calculated 
transverse mass spectrum.
For the kaons, one observes that the measured transverse mass spectra exhibit a
convex curvature indicative for collective flow of matter.
This is a nontrivial dynamical effect not contained in our
simple phase-space approach. Collective flow is more prominent
for heavier particle species not considered here and will lead
to similar deviations from the spectra calculated within our simple
phase-space approach. Nevertheless,
we expect that the quality of the fit can be further improved by
introducing separate parameters $a_r$ for each particle species $r$.
This, however, as well as a more refined treatment of resonance
decays, is beyond the scope of the present work where we
want to keep the treatment as simple as possible.

\section{Summary and Outlook} \label{V}

We have studied particle production in nuclear collisions
within a phase-space model. 
Assuming a simple form for the matrix element
that describes the production of $n$ secondary particles, we
have computed the phase-space integral within a statistical approach based
on the Laplace transformation and the saddle-point method.
Single-inclusive invariant momentum spectra
can be then computed from the ratio of the phase-space integrals
for $n-1$ and $n$ particles, respectively. 

We have applied this phase-space approach to multiparticle
production to data for Pb+Pb collisions at 158 AGeV taken by the 
NA49 collaboration. In order to compute the phase-space integral,
we require (a) the total available energy which is determined
from the beam energy and the number of wounded nucleons 
(in our case, $\sqrt{s}= 3131.3$ GeV) and (b)
the multiplicities and the masses 
of the various particle species produced in the reaction.
However, the NA49 experiment does not measure all
species of strongly interacting particles.
We reconstructed the multiplicities of the most important missing particle
species using symmetry considerations, 
conservation laws, a simple quark model and an empirical rule.
In this way we obtained 2611 secondary particles.
By changing this value by - 10\% and +20\%, respectively, 
we tested the sensitivity
of our results to the reconstruction procedure and found
that this does not affect our conclusions.

{}From the phase-space integral we then computed
transverse mass and rapidity spectra for $\pi^-,\, K^+$, and $K^-$
mesons and compared them with the measured spectra.
Our simple model predicts transverse mass spectra of exponential
shape, with an inverse slope parameter that is (almost)
independent of the particle mass. 

We also computed transverse mass and rapidity spectra
assuming the system to be in thermal equilibrium at a temperature
which is determined from the total available energy $\sqrt{s}
=3131.3$ GeV. In the thermal model, transverse mass spectra
are not pure exponentials, but are of slightly convex shape.
At midrapidity, the inverse slope parameter of the transverse mass 
spectra is identical to the temperature, $T$, and smaller than
the inverse slope parameter, $T^*$, arising in the phase-space model.
This difference can be traced to the different definitions of
phase space in the two models, invariant phase space 
$d^3{\bf p}_i/E_i$ for the phase-space model, and
non-invariant phase space $d^3{\bf p}_i$ for the thermal model.
For ultrarelativistic particles, the difference in
the slope parameters can be computed analytically,
$T^*/T = 3/2$.
Due to the convex shape of the transverse mass spectra in the
thermal model, in the range of transverse masses $0 \leq m_\perp -m
\leq 1$ GeV, the transverse mass spectra in both models are
surprisingly similar. The difference in the slope parameters 
becomes visible only at larger transverse masses.
The rapidity spectra of the thermal and the phase-space model 
are also very similar. 

If we assume a constant, i.e., isotropic matrix element for multiparticle
production, the transverse mass and rapidity spectra
calculated from the phase-space model
deviate substantially from the experimentally measured
spectra. The same holds for the thermal model.
This deviation is due to the fact that there is
still a remnant of the initial, longitudinal motion 
of the nuclei.

There are two possibilities to obtain agreement between models
and data. The first possibility is to retain the assumption
of an isotropic phase-space distribution {\em locally\/}
in coordinate space. One then employs
a dynamical model for the evolution of matter, such as
fluid dynamics, which gives rise to collective transverse as well as
longitudinal flow of matter. For a given equation of state,
one can then tune the initial as well as the freeze-out
conditions in order to reproduce the data and thus account
for longitudinal phase-space dominance.

In this paper, we pursue the second possibility. We
relax the assumption of isotropic phase space and take
longitudinal phase-space dominance into account by
a simple ansatz for the matrix element which exponentially
suppresses transverse momenta. This has the effect
that the transverse mass spectra decrease faster 
and the rapidity spectra become wider.
The amount of suppression is determined by a parameter $a$
which is fitted to the experimental spectra.
The best fits were obtained by 
choosing different parameters $a$ for each type of particle 
($\pi^+,\, K^+,\, K^-$)
and fitting transverse mass spectra and rapidity spectra separately.
Surprisingly, a rather good fit of transverse mass and
rapidity spectra of all particles considered can also be obtained 
by a single value, $a = (0.256\, {\rm GeV})^{-1}$.
Note that this second approach does not require 
the assumption of thermal equilibrium.
Observed deviations between calculated and measured spectra
can be attributed to nontrivial dynamical phenomena, such
as resonance decays and collective flow.

A possible extension of our present work is to 
compare our phase-space model with data for heavier
particles, such as nucleons and hyperons. 
It is known that the transverse mass spectra of
particles heavier than pions and kaons have a substantially
larger inverse slope parameter (but essentially of similar magnitude for
all of them). This is usually attributed
to the collective radial flow of matter \cite{NuXu}.
In order to incorporate this into our approach, we would 
have to introduce one additional
parameter $a$ for the heavier particle species.
The other way to account for different slope parameters for
the various particle species would be to
use a more physical ansatz for, 
or even a detailed calculation of,
the matrix element $T({\bf p}_i)$ \cite{hsu}.
Another possible extension is to incorporate the distortion of the
phase-space of light particles, such as pions, by the decay
of heavier resonances \cite{Byck}. 

\section*{Acknowledgment}

We thank A.\ Andronic, F.\ Becattini, M.\ Bleicher, A.\ Dumitru, C.\ Greiner, 
M.\ Gyulassy, K.\ Kajantie,
J.\ Knoll, A.\ Kostyuk, J.\ Reinhardt, J.\ Schaffner-Bielich, 
I.\ Shovkovy, M.\ Strikman, Q.\ Wang, 
and G.\ Wilk for comments and discussions.
We thank in particular M.\ Ga\'{z}dzicki for valuable discussions 
concerning the reconstruction of the multiplicities of 
particles in the experiment and for providing us with the
multiplicities given in Table \ref{tab2}.
We also thank J.\ Stachel and P.\ Braun-Munzinger for
providing us with a numerical routine for the saddle-point method.


\begin{thebibliography}{70}
\bibitem{QM2004} see, for instance, proc.\ of the 17th int.\ conf.
on ultra-relativistic nucleus-nucleus collisions ``Quark Matter
2004'', Oakland, USA, Jan.\ 11\-17, 2004 (eds.\
H.G.\ Ritter, X.N.\ Wang), J.\ Phys.\ G30 (2004) 633.
\bibitem{latt} for a review see, for instance, E.\ Laermann and
O.\ Philipsen, Ann.\ Rev.\ Nucl.\ Part.\ Sci.\ 53 (2003) 163.
\bibitem{huovinen} P.\ Huovinen in {\it Quark-gluon plasma}
(eds.\ R.C.\ Hwa, X.N.\ Wang, World Scientific, Singapore, 2004) 600.
\bibitem{sQGP} T.D.\ Lee, Nucl.\ Phys.\ A750 (2005) 1;
M.\ Gyulassy and L.\ McLerran, {\it ibid.} 30;
E.\ Shuryak, {\it ibid.} 64.
\bibitem{kolb} P.F.\ Kolb and U.W.\ Heinz in {\it Quark-gluon plasma}
(eds.\ R.C.\ Hwa, X.N.\ Wang, World Scientific, Singapore, 2004) 634.
\bibitem{Braun} P.\ Braun-Munzinger, K.\ Redlich, and J.\ Stachel
in {\it Quark-gluon plasma}
(eds.\ R.C.\ Hwa, X.N.\ Wang, World Scientific, Singapore, 2004) 491.
\bibitem{Becca} F.\ Becattini, M.\ Ga\'{z}dzicki, A.\ Keranen, J.\ 
Manninen, R.\ Stock, Phys.\ Rev.\ C69 (2004) 024905.
\bibitem{Cley} S.\ Wheaton and J.\ Cleymans, hep-ph/0407174.
\bibitem{hirano} T.\ Hirano and K.\ Tsuda, Phys.\ Rev.\ C66 (2002) 054905.
\bibitem{gyul} M.\ Gyulassy and T.\ Hirano, nucl-th/0506049.
\bibitem{rasanen} K.J.\ Eskola, H.\ Niemi, P.V.\ Ruuskanen, and S.S.\
Rasanen, Nucl.\ Phys.\ A715 (2003) 561c.
\bibitem{Hag} R.\ Hagedorn, {\it Relativistic Kinematics}
(W.A.\ Benjamin, 1964).
\bibitem{Byck} E.\ Byckling and K.\ Kajantie, 
{\it Particle Kinematics} (John Wiley \& Sons, 1973).
\bibitem{Hama} Y.\ Hama and M.\ Pl\"umer, Phys.\ Rev.\ D 46 (1992) 160.
\bibitem{knoll} J.\ Knoll, Phys.\ Rev.\ C20 (1979) 773;
S.\ Bohrmann and J.\ Knoll, Nucl.\ Phys.\ A356 (1981) 498.
\bibitem{Risch} D.H.\ Rischke, Nucl.\ Phys.\ A698 (2002) 153c.
\bibitem{hsu} J.\ Hormuzdiar, S.D.H.\ Hsu, and G.\ Mahlon,
Int.\ J.\ Mod.\ Phys.\ E12 (2003) 649.
\bibitem{Koch} V.\ Koch, Nucl.\ Phys.\ A715 (2004) 108c.
\bibitem{Becca2} F.\ Becattini, J.\ Phys.\ Conf.\ Ser.\ 5 (2005) 175.
\bibitem{NA49} Marek Ga\'{z}dzicki, private communication.
\bibitem{Afan} S.V.\ Afanasiev et al., Phys.\ Rev.\ C66 (2002) 054902.
\bibitem{Misch} A.\ Mischke et al., Nucl.\ Phys.\ A715 (2003) 453.
\bibitem{NA49lam} T.\ Anticic et al., Phys.\ Rev.\ Lett.\ 93 (2004) 022302.
\bibitem{NA49xi} S.V.\ Afanasiev et al., Phys.\ Lett.\ B538 (2002) 275.
\bibitem{NA49ome} C.\ Alt et al., Phys.\ Rev.\ Lett.\ 94 (2005) 192301.
\bibitem{werner} K.\ Werner and J.\ Aichelin, Phys.\ Rev.\ C52 (1995)
1584.
\bibitem{Fri} B.L.\ Friman and J.\ Knoll, {\it Introduction to the
physics of hot and dense hadronic matter}, lectures presented at GSI
summer school 2000.
\bibitem{Jam} F.\ James, {\it Monte Carlo Phase Space}, CERN Report 68-15.
\bibitem{Kleiss} R.\ Kleiss and W.J.\ Stirling, Nucl.\ Phys.\ B385 (1992)
413.
\bibitem{Lurc} F.\ Lur\c{c}at, P.\ Mazur, Nuov.\ Cim.\ XXXI (1964) 140.
\bibitem{Krzy} A.\ Krzywicki, Nuov.\ Cim.\ XXXII (1967) 1067.
\bibitem{Krzy1} A.\ Krzywicki, J.\ Math.\ Phys.\ 6 (1965) 485.
\bibitem{Bial} A.\ Bialas and T.\ Ruijgrok, Nuov.\ Cim.\ XXXIX (1965) 1061.
\bibitem{Kaja} K.\ Kajantie and V.\ Karim\"aki, Comp.\ Phys.\ Commun.\ 2
(1971) 207.
\bibitem{Kaja1} K.\ Kajantie and V.\ Karim\"aki,  Ann.\ Acad.\ Sci.\
Fenn.\ Series A, VI.\ Physica, Vol.\ 395 (Helsinki, 1972)
\bibitem{Khin} A.J.\ Khinchin, {\it Mathematical Foundations
of Statistical Mechanics} (Dover, New York, 1949).
\bibitem{Grad} I.S.\ Gradsteyn and I.M.\ Ryshik, {\it Tables of
Integrals, Sums, Series and Products}, Eq.\ (8.432.9)
(Verlag Harri Deutsch, Frankfurt, 1981).
\bibitem{Abra} M.\ Abramowitz and I.A.\ Stegun, 
{\it Handbook of Mathematical Functions} (Dover, New York, 1965).
\bibitem{Coop} F.\ Cooper, G.\ Frye, and E.\ Schonberg, Phys.\ Rev.\ D11
(1975) 192.
\bibitem{Wrob} A.\ Wroblewski, Acta Phys.\ Pol.\ B16 (1985) 379.
\bibitem{Gazdz} M.\ Gazdzicki and O.\ Hansen, Nucl.\ Phys.\ A528 (1991) 754.
\bibitem{Raf} J.\ Rafelski, Phys.\ Lett.\ B262 (1991) 333.
\bibitem{Bial1} A.\ Bialas, Phys.\ Lett.\ B442 (1998) 449.
\bibitem{NuXu} Nu Xu for the NA44 coll., Nucl.\ Phys.\ A610 (1996) 175c.
\end{thebibliography}
\end{document}